\documentclass[useAMS,usenatbib]{mn2e}

\usepackage{psfig}
\usepackage{graphicx}
\usepackage{times}
\usepackage{natbib}

\newif\ifAMStwofonts
\AMStwofontstrue

\newcommand{\be}{\begin{equation}}
\newcommand{\ee}{\end{equation}}
\newcommand{\ba}{\begin{eqnarray}}
\newcommand{\ea}{\end{eqnarray}}
\newcommand{\brr}{\begin{array}}
\newcommand{\err}{\end{array}}
\newcommand{\bc}{\begin{center}}
\newcommand{\ec}{\end{center}}
\newcommand{\hm}{\,h^{-1}{\rm Mpc}}

\newcommand{\msun}{\,h^{-1}M_\odot}

\newcommand{\tmw}{\mbox{$T_{\rmn{mw}}$}}

\newcommand{\tsz}{tSZ}

\newcommand{\tsl}{\mbox{$T_{\rm {sl}}$}}

\newcommand{\fwid}{7.cm}

\newcommand{\mincir}{\raise
  -2.truept\hbox{\rlap{\hbox{$\sim$}}\raise5.truept \hbox{$<$}\ }}
\newcommand{\magcir}{\raise
  -2.truept\hbox{\rlap{\hbox{$\sim$}}\raise5.truept \hbox{$>$}\ }}
\newcommand{\siml}{\raise
  -2.truept\hbox{\rlap{\hbox{$\sim$}}\raise5.truept \hbox{$<$}\ }}
\newcommand{\simg}{\raise
  -2.truept\hbox{\rlap{\hbox{$\sim$}}\raise5.truept \hbox{$>$}\ }}

\newcommand{\eq}[1]{eq.~(\ref{#1})}

\newcommand{\fig}[1]{Figure~\ref{#1}}

\newcommand{\figs}[2]{Figures~\ref{#1} and \ref{#2}}
\newcommand{\figss}[3]{Figures~\ref{#1}, \ref{#2} and \ref{#3}}
\newcommand{\tab}[1]{Table~\ref{#1}}

\title[Mass profiles from SZ/X--ray in simulated clusters]
      {Reconstructing mass profiles of simulated galaxy clusters by
        combining Sunyaev--Zeldovich and X--ray images}
\author[Ameglio et al.]  {S. Ameglio$^{1,2,3,4}$,
S. Borgani$^{1,2,3}$, E. Pierpaoli$^4$, K. Dolag$^5$, S. Ettori$^{6,7}$ \&
A. Morandi$^8$\\~\\ $^1$
Dipartimento di Astronomia dell'Universit\`a di Trieste, via Tiepolo
11, I-34131 Trieste, Italy (borgani@oats.inaf.it) \\ $^2$ INFN
-- National Institute for Nuclear Physics, Trieste, Italy\\ $^3$ INAF
-- Osservatorio Astronomico di Trieste, Trieste, Italy\\ $^4$
University of Southern California, Los Angeles, CA
(ameglio,pierpaol@usc.edu)\\ $^5$ Max-Planck-Institut f\"ur Astrophysik,
Karl-Schwarzschild Strasse 1, Garching bei M\"unchen, Germany
(kdolag@mpa-garching.mpg.de)\\ $^6$ INAF -- Osservatorio Astronomico
di Bologna, via Ranzani 1, I-40127 Bologna, Italy
(ettori@oabo.inaf.it)\\ $^7$ INFN, Sezione di Bologna, viale Berti Pichat 6/2, I-40127 Bologna, Italy\\$^8$ Dipartimento di Astronomia, Universit\`a
di Bologna, via Ranzani 1, I-40127 Bologna, Italy (morandi@oabo.inaf.it) }

\begin{document}

\date{Accepted ???. Received ???; in original form ???}

\maketitle

\begin{abstract}
  We present a method to recover mass profiles of galaxy clusters by
  combining data on thermal Sunyaev--Zeldovich (\tsz) and X--ray
  imaging, thereby avoiding to use any information on X--ray
  spectroscopy. This method, which represents a development of the
  geometrical deprojection technique presented in
  \cite{2007MNRAS.382..397A}, implements the solution of the
  hydrostatic equilibrium equation. In order to quantify the
  efficiency of our mass reconstructions, we apply our technique to a
  set of hydrodynamical simulations of galaxy clusters.  We propose
  two versions of our method of mass reconstruction. Method 1 is
  completely model--independent and assumes as fitting parameters the
  values of gas density and total mass within different radial bins.
  Method 2 assumes instead the analytic mass profile proposed by
  \cite{NA97.1} (NFW). We find that the main source of bias in
  recovering the mass profiles is due to deviations from hydrostatic
  equilibrium, which cause an underestimate of the mass of about 10
  per cent at $r_{500}$ and up to 20 per cent at the virial radius.
  Method 1 provides a reconstructed mass which is biased low by about
  10 per cent, with a 20 per cent scatter, with respect to the true
  mass profiles.  Method 2 proves to be more stable, reducing the
  scatter to 10 per cent, but with a larger bias of 20 per cent,
  mainly induced by the deviations from equilibrium in the
  outskirts. To better understand the results of Method 2, we check
  how well it allows to recover the relation between mass and
  concentration parameter.  When analyzing the 3D mass profiles we
  find that including in the fit the inner 5 per cent of the virial
  radius biases high the halo concentration, thus suggesting that the
  NFW profile is not a perfect fit in the central regions of our
  simulations including cooling and star formation. Also, at a fixed
  mass, hotter clusters tend to have larger concentration. Our
  procedure recovers the concentration parameter essentially unbiased
  but with a scatter of about 50 per cent. In general, our analysis
  demonstrates that combining X--ray imaging with spatially resolved
  \tsz\ data is a valid alternative to using X--ray spectroscopy to
  recover the mass of galaxy clusters.
\end{abstract}


\section{Introduction} 
A number of important cosmological tests are based on mass
measurements in galaxy clusters. In particular the mass function, the
baryon fraction and their redshift evolution are highly sensitive to
the underlying cosmology and provide constraints on the Dark
  Matter and Dark Energy content of the Universe. Precise mass
measurement in galaxy cluster are then necessary to calibrate clusters
as precision tools for cosmology \citep{2001ApJ...553..545H,
  2002ARA&A..40..539R, 2003MNRAS.342..163P, 2005RvMP...77..207V,
  2006astro.ph..5575B}.

In X--ray studies, the total collapsed mass of a cluster is determined
by applying the hydrostatic equilibrium equation to gas density and
temperature profiles. A number of authors
\citep[e.g.,][]{2004MNRAS.351..237R,2004MNRAS.355.1091K,2007ApJ...655...98N}
analyzed hydrodynamical simulations of galaxy clusters and found that
the gas is not in perfect hydrostatic equilibrium. Instead, they found
deviations up to 20 per cent, which have the effect of systematically
biasing low the reconstruction of the total collapsed mass.  The
amount of this underestimate depends on both the model assumed and the
dynamical state of the cluster.  For instance,
\cite{2006MNRAS.369.2013R} showed that assuming the isothermal
$\beta$--model \citep{1976A&A....49..137C} for the gas distribution
 gives the worst reconstruction. \cite{2007ApJ...655...98N}
differentiated their sample of simulated clusters in relaxed and
unrelaxed objects, with the latter showing a larger scatter in the
mass reconstruction. Following the same direction,
\cite{2007arXiv0708.1518J} found a correlation between quantitative
measures of the morphology of the X--ray images and the bias in the
mass reconstruction, although with a quite large scatter.
\cite{2007A&A...474..745P} probed the deviations from hydrostatic
equilibrium in different stages of a merger process, while
\cite{2008arXiv0808.1111P} performed an extended analysis of a large
sample ($\sim 100$) of simulated galaxy clusters, in order to
disentangle various biases in the mass reconstruction.

Recovering mass profiles via the hydrostatic equilibrium equation
involves the derivatives of both gas density and temperature
profiles. Then, an accurate mass determination requires high--quality
temperature measurements. For this reason, X--ray studies are often
limited to the inner regions of the clusters and to objects at
moderate redshift.  Furthermore, all the temperature profiles so far
used for the reconstruction of cluster masses are based on X--ray
data. As originally noticed by \cite{2004MNRAS.354...10M} \citep[see
also][]{2006ApJ...640..710V}, the thermal complexity of the ICM causes
the temperature determined by fitting the X--ray spectrum to a
single--temperature bremsstrahlung model to be generally different
from the electron temperature. On the other hand, it is the electron
temperature that enters in the equation of hydrostatic equilibrium
(under the assumptions of fully ionized plasma and equilibration
between ions and electrons). The question then arises as to whether
the difference between spectroscopic and electron temperature may
induce an additional bias in the mass estimate.

In this paper, we propose to use a combination of X--ray imaging and
\tsz\ data, which avoids the use of X--ray spectroscopy.  Taking
advantage of the different dependence of the \tsz\ signal and of the
X--ray emissivity on gas density and temperature, one can recover both
by suitably deprojecting \tsz\ and X--ray imaging data, without the
need of any X--ray spectroscopic information. Furthermore, both X--ray
and SZ cluster images can be obtained in principle out to larger radii
than possible for X--ray spectra. Therefore, their combination could
allow studies of the ICM thermal properties to be pushed out to larger
fractions of the whole cluster virialized regions.  Indeed, different
algorithms, which use a combination of \tsz\ and X--ray data, have
been proposed by a number of authors and applied to analytical cluster
models and/or sets of simulated clusters \citep[e.g.][and references
  therein]{2001ApJ...561..600Z, 2004ApJ...601..599L,
  2005ApJ...625..108D, 2007A&A...474..745P, 2007MNRAS.382..397A}.
Owing to the relatively low angular resolution of currently available
\tsz\ telescopes, this method has been so far applied only to a
handful of observed clusters \citep{2000ApJ...545..141Z,
    2002A&A...387...56P, 2008arXiv0809.5077M}, with results that are
generally consistent with those based on X--ray spectroscopic data.

Accurate mass profiles reconstruction can also be used as probes for
cosmology. In fact, mass profiles are expected to follow a universal
functional form, which is valid over a wide range of masses, from 
  dwarf galaxies to massive galaxy clusters. A formulation for this
function has been originally provided by fitting the mass profiles in
a set of N--body simulations by \cite{NA97.1} (NFW hereafter). The
analysis of X--ray data from Chandra and XMM--Newton observations
confirm the validity of the NFW model, out to a significant fraction
of the cluster virial radius \citep[e.g.][]{2002A&A...394..375P,
  2005A&A...435....1P, 2006ApJ...640..691V, 2006ApJ...650..777Z}. An
important parameter of the NFW model is the concentration $c$, which
is given by the ratio between the halo's virial radius and the
characteristic radius of the density profile.  In the hierarchical
structure formation scenario the more massive objects are expected to
form recently, from a lower--density environment than less massive
ones. Thus, it is expected an inverse correlation between the mass of
an object and its concentration, with a substantial scatter related to
the distribution of halo formation epochs
\citep[e.g.,][]{NA97.1,BU01.1,2004A&A...416..853D}. This relation
  has been confirmed by the observation of mass profiles in galaxy
  clusters, using both X-ray and lensing data \citep[][and references
    therein]{2007MNRAS.379..209S,2008JCAP...08..006M}. These authors
  generally agree in finding a well defined relation with a
  substantial intrinsic scatter, as expected from the predictions
  of numerical simulations. However, some discrepancies are
  still present in the determination of the slope and normalization of
  the relation, which may be generated by biases in the mass
  measurements and/or by selection effects.

We extend here the deprojection algorithm presented by
\cite{2007MNRAS.382..397A} (A07 hereafter), by including the solution of the
hydrostatic equilibrium equation.  Using this technique, we analyze a
set of 14 simulated clusters having $\tsl \magcir 3$~keV, with the aim
of quantifying the accuracy with which total mass profiles can be
recovered by combining X--ray and \tsz\ images. We will
also discuss how the relation between halo concentration $c$ and mass
$M$ can be recovered with our deprojection method, also comparing it
with the theoretical predictions of the model by
\cite{2001ApJ...554..114E}.  The analysis presented in the following
can be applied to data from the present
generation of X--ray telescopes. Furthermore, future X--ray
observations with lower background (e.g. as expected from the eROSITA
mission\footnote{http://www.mpe.mpg.de/projects.html\#erosita}) will
provide good imaging data for a large number of clusters out to
$z\simeq 1$. As for the tSZ data, exploiting the full potentiality of
our technique would require spatially resolved observations, which
will be available from upcoming (or just started) SZ experiments,
based both on interferometric arrays (ALMA: Atacama Large Millimeter
Array\footnote{http://www.eso.org/projects/alma/}; CARMA: Combined
Array for Research in Millimeter-wave
Astronomy\footnote{http://www.mmarray.org}) and on single dishes with
large bolometer arrays (CCAT: Cornell--Caltech Atacama Telescope
\footnote{http://astrosun2.astro.cornell.edu/research/projects/atacama/};
LMT: Large Millimeter Telescope\footnote{http://www.lmtgtm.org/}).

The paper is structured as follows. In Section \ref{sec:sim} we
briefly present the set of simulated clusters. We use the same subset
of simulated clusters having $T_{sl} \magcir 3$ keV, which is
described by A07.  In Section \ref{sec:hyd-equil} we introduce the
hydrostatic equilibrium equation and, as a preliminary test, probe the
intrinsic deviations from this equilibrium for our set of simulated
clusters. After briefly reviewing the deprojection algorithm
introduced by A07, Section \ref{sec:total-method} describes the
implementation of the hydrostatic equilibrium equation in this
algorithm. Sections \ref{sec:total-results} and \ref{sec:cm-rel}
present our results on the total mass reconstruction and on the
$c$--$M$ relation respectively. Finally, our main conclusions are
summarized in Section \ref{sec:total-concl}.

\section{The set of simulated clusters}\label{sec:sim} 
The sample of simulated galaxy clusters used in this paper has been
extracted from a large-scale cosmological hydro-N-body simulation of a
``concordance'' $\Lambda$CDM model with $\Omega_m=0.3$ for the matter
density parameter, $\Omega_\Lambda=0.7$ for the cosmological constant,
$\Omega_{\rm b}=0.019\,h^{-2}$ for the baryon density parameter,
$h=0.7$ for the Hubble constant in units of 100 km s$^{-1}$Mpc$^{-1}$
and $\sigma_8=0.8$ for the r.m.s. density perturbation within a
top--hat sphere having comoving radius of $8\hm$ (see
\citealt{2004MNRAS.348.1078B}, for further details).  The simulation,
performed with the Tree+SPH code {\small GADGET-2}
\citep{2005astro.ph..5010S}, follows the evolution of $480^3$ dark
matter particles and an initially equal number of gas particles in a
periodic cube of size $192 h^{-1}$ Mpc. The mass of the gas particles
is $m_{\rm gas}=6.9 \times 10^8 h^{-1} M_\odot$, and the
Plummer-equivalent force softening is $7.5 h^{-1}$ kpc at $z=0$.  The
simulation includes the treatment of radiative cooling, a uniform
time--dependent UV background, a sub--resolution model for star
formation and energy feedback from galactic winds
\citep{2003MNRAS.339..289S}. At $z=0$ we extract a set of 117
clusters, whose mass, as computed from a friends-of-friends algorithm
with linking length $b=0.15$ (in units of the mean interparticle
distance) is larger than $10^{14}\msun$.  For these clusters we
compute the spectroscopic--like temperature
\be
T_{sl}=\frac{\sum_i n_{e,i}^2 T_i^{a-1/2}}{\sum_i n_{e,i}^2 T_i^{a-3/2}}\,,
\label{eq:tsl}
\ee
where $a=0.75$ is a parameter shown by
\cite{2004MNRAS.354...10M} to accurately reproduce the temperatures
obtained from the spectroscopic fit.

Due to the limited box size, the largest cluster found in the
cosmological simulation has $\tsl=4.6$ keV.  In order to extend our
analysis to more massive and hotter systems, which are mostly relevant
for current tSZ observations, we include four more galaxy clusters
having $M_{\rm vir}>10^{15} \msun$\footnote{Here and in the following,
  the virial radius, $r_{\rm vir}$, is defined as the radius of a
  sphere centred on the local minimum of the potential, containing an
  average density, $\rho_{\rm vir}$, equal to that predicted by the
  spherical collapse model. For the cosmology assumed in our
  simulations at low redshift it is $\rho_{\rm vir}\simeq 100
  \rho_{\rm c}$, being $\rho_{\rm c}$ the cosmic critical
  density. Accordingly, the virial mass, $M_{\rm vir}$, is defined as
  the total mass contained within this sphere.}  and belonging to a
different set of hydro-N-body simulations
\citep{2006MNRAS.tmp..270B}. These objects have been selected from a
DM--only simulation of a large cosmological volume \citep{YO01.1}, and
resimulated at higher resolution. The achieved resolution corresponds
to $m_{\rm gas}= 1.69 \times 10^{8} h^{-1}M_\odot$ for the mass of the
gas particle and a gravitational softening of $5 h^{-1}$kpc at
$z=0$. These simulations have been performed with the same choice of
the parameters defining star--formation and feedback. The cosmological
parameters also are the same, except for a higher power spectrum
normalization, $\sigma_8=0.9$.

\begin{table}
\centerline{
\begin{tabular}{lcccc}
Cluster  & $M_{vir}$& $T_{sl}$ & $r_{vir}$& $r_{500}$ \\
        & $10^{14}M_{\odot}$ & keV & Mpc & Mpc\\
C1 & 5.4  & 3.1  & 2.1  & 1.0  \\
C2 & 10.1 & 4.3  & 2.6  & 1.3  \\
C3 & 18.6 & 4.6  & 3.2  & 1.5  \\
C4 & 21.4 & 6.8  & 3.3  & 1.7  \\
C5 & 9.9  & 4.5  & 2.6  & 1.3  \\
C6 & 5.7  & 3.1  & 2.1  & 1.0  \\
C7 & 7.1  & 3.6  & 2.3  & 1.1  \\
C8 & 5.8  & 3.2  & 2.2  & 1.1  \\
C9 & 4.8  & 3.0  & 2.0  & 1.0  \\
C10 & 4.8  & 3.0  & 2.0  & 1.0  \\
C11 & 6.3  & 3.0  & 2.2  & 1.0  \\
C12 & 32.0 & 8.9  & 3.8  & 1.9  \\
C13 & 19.2 & 6.3  & 3.2  & 1.6  \\
C14 & 19.4 & 5.7  & 3.2  & 1.6  \\
\end{tabular}
}
\caption{Characteristics of the simulated clusters, having $\tsl
  \magcir 3$~keV, to which we apply the deprojection
  procedure. Col. 2: virial mass; Col. 3: spectroscopic--like
  temperature; Col. 4: virial radius; Col. 5: $r_{500}$.}
\label{tab:clusters}
\end{table}

From this large set of simulated clusters, we select a subsample of
objects having $T_{sl} \magcir 3$~keV, which are more relevant for SZ
studies. We identify 15 suitable objects, from which we exclude one
very irregular cluster. The main characteristics of this subset of
simulated clusters are listed in \tab{tab:clusters}, where the cluster
labeled C4, C12, C13 and C14 are those based on the high--resolution
resimulations. All the results shown in this paper are based on the
analysis of these 14 objects. A larger subsample of systems with
$T_{sl}< 3$ keV, extracted from the same box, are used for the
analysis presented in Section 6.

We generate synthetic X--ray and tSZ maps of these clusters, to which
we apply our method of mass reconstruction. Our procedure of map
making is described in detail in A07, while here we summarize the most
important characteristics: {\it i}) we generate three maps of each
cluster, by projecting it on the principal axes of the inertia tensor,
so that the projection on the $x$, $y$ and $z$ axes represent the
projection along the minimum, medium and maximum elongation of the
cluster, respectively; {\it ii}) the X--ray maps are obtained by
Montecarlo generation of $10^4$ photon counts from the surface
brightness maps in the [0.5-2] keV energy band, after smoothing it
with a PSF having a FWHM of 0.5 arcsec (comparable to the Chandra one
at the aim point), without adding the contribution from any
instrumental or cosmic background; {\it iii}) the noise setup for the
tSZ maps is modeled on the Cornell Caltech Atacama Telescope
(CCAT). The maps are convolved with a Gaussian beam having
FWHM$=0.44'$ and then it is added a white noise.  In order to have
approximately the same signal--to--noise for all objects, the noise
level is 3 $\mu$K/beam for clusters with $\tsl < 4$~keV and
10~$\mu$K/beam for the hotter ones.

In the following, we will show detailed results on the recovery of
total mass profiles for a subset of 4
clusters, which are indicated in \tab{tab:clusters} with
C1--4. The first three of them are extracted from the cosmological
hydrodynamical simulation, while C4 is one of the massive clusters
simulated at higher resolution.  C2, C3 and C4 are typical examples of
clusters at low, intermediate and high temperature, while C1 is an
interesting case to understand the effect of fore--background
contaminations. These four clusters have been used also in A07
as examples.

\section{The hydrostatic equilibrium} \label{sec:hyd-equil} 
Total mass profiles from observations of the ICM are computed by
assuming that the gas lies in hydrostatic equilibrium (HE hereafter)
within the cluster gravitational potential. In the case of spherical
symmetry, the equation of hydrostatic equilibrium can be cast in the
form
\be\label{eq:hydeq}
  M_{tot}(<r) = - {kT(r)\;r \over G \mu m_p} \left[ {d\ln n_e
  \over d \ln r } +{d\ln T \over d \ln r } \right]\,,
\ee
where $T(r)$ is the temperature at the radius $r$, $\mu$ the mean
molecular weight ($\mu \simeq 0.6$ for a gas of primordial
composition), and $m_p$ the proton mass. Note that the mass at a given
radius depends only upon the local pressure derivative and is
unaffected by the physical properties of the cluster at smaller or
larger radii.

While to first approximation clusters are quite close to the condition
of pressure equilibrium, small but sizable deviations are generally
found in the analysis of simulated clusters 
\citep[e.g.,][]{2004MNRAS.351..237R, 2004MNRAS.355.1091K,
2004ApJ...606L..97D, 2006MNRAS.369.2013R, 2007ApJ...655...98N,
2008arXiv0808.1111P}. These deviations are generally ascribed to the
presence of non--negligible stochastic gas motions,
which provide an effective non--thermal pressure support, thereby
leading to an underestimate of the total gravitating mass when not
accounted for in \eq{eq:hydeq}.  \cite{2004MNRAS.351..237R}
suggested the addition of an extra--term, which takes into account the
presence of gas motions, to the equation of the hydrostatic
equilibrium. \cite{2007arXiv0708.1518J} find a correlation between the
amount of substructures and the underestimate of the total mass, in a
set of hydrodynamical simulations. However, the large scatter in this
correlation around the mean relation suggests that substructures may
not be the only sources of the bias in the mass reconstruction.

\begin{figure}
\centerline{
\psfig{file=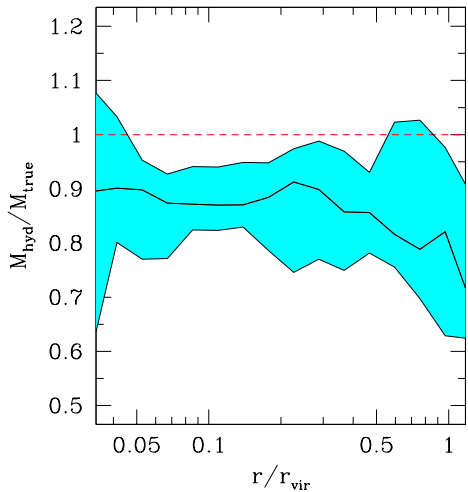,width=\fwid}
}
\caption{The ratio between the hydrostatic and the true mass profiles,
  averaged over the set of simulated clusters. The shaded area
  encompasses the 16 and 84 percentiles, while the central continuous
  line shows the average profile of the deviation.}
\label{fi:mhyd}
\end{figure}

Before applying our procedure of mass reconstruction, we assess the
degree of violation of HE in our clusters. This will allow us to
quantify by how much the differences between reconstructed and true
mass profiles are due to biases in the deprojection method or to
violation of the HE.

We apply \eq{eq:hydeq} to the true, 3-D density and temperature
profiles, by performing a numerical derivative in the log-log space,
using a 3--point Lagrangian interpolation\footnote{The profile of
  $\log(nT)$ is computed for log-equispaced values of $r$. To compute
  the derivative at a given radius, we fit with the point of interest
  and its two adjacent ones with a quadratic curve. The first-order
  derivative at the middle point is taken to be the derivative of the
  profile at that radius.}.  We will refer to the mass so computed as
the hydrostatic mass, $M_{hyd}$ hereafter.  In \fig{fi:mhyd} we
show the profiles of the ratio $M_{hyd}/M_{true}$ averaged over our
set of simulated clusters, along with the $1\sigma$ scatter around the
mean.  We generally find that the hydrostatic masses underestimate the
true ones on average by $\sim$ 10 per cent out to about $r_{500}$, a
result which is in line with those found by other analyses of
simulated clusters \citep[e.g.,][]{2007ApJ...655...98N,
  2007A&A...474..745P, 2007arXiv0708.1518J}.  At larger radii this
underestimate increases, reaching about 20 per cent at $r_{vir}$,
along with an increase of the scatter. This is consistent with the
expectation that outer cluster regions deviate more from the condition
of HE, due to the presence of ongoing mergers and continuous gas
accretion.

\section{Methods of mass profile reconstruction} 
\label{sec:total-method} 
The methods described here, aimed at recovering the cluster mass
profiles, represent a development of the MonteCarlo Markov Chain
(MCMC) maximum likelihood deprojection technique described in A07. The
algorithm is modified in order to solve the hydrostatic equilibrium
equation while deprojecting the cluster images. In this way, the
temperature profile is computed from the gas density and total mass
profiles, so that all quantities are derived simultaneously and in a
fully self--consistent way.  We provide here below only a brief
description of the deprojection technique, while we refer to A07 for
further details.

The cluster is assumed to have a onion-skin structure, with N
concentric spherical shells, having uniform gas properties (gas
density and temperature). The image of the cluster (in both tSZ and
X--ray) is divided into $N$ rings, which have the same limiting radii
of the shells for sake of simplicity. The \tsz\ and X--ray profiles
are then computed by projecting this onion-skin model in the plane of
the sky. We divide the virial radius and $r_{500}$ into $N=14$ and
$N=10$ rings, respectively, equally spaced in logarithm.  This
solution represents a good compromise between the resolution of the
profiles and the noise level.  Finally, the inner 50 kpc of all
clusters are excluded by all fits.

The deprojection is then performed by the maximization of a likelihood
function, which is computed by comparing the observed \tsz\ and X--ray
surface brightness profiles with the ones obtained from the
model. This approach has the advantage of deprojecting of both X--ray
and \tsz\ profiles simultaneously, directly obtaining the whole gas
density and temperature and total mass profiles, along with the
corresponding uncertainties. Moreover, it is possible to introduce in
the likelihood extra terms in order to improve the accuracy and
robustness of the technique. In particular, we adopt a regularization
constraint, based on the Philips-Towmey regularization method
\citep[][and references therein]{1995A&AS..113..167B}. This method has
been already used also by \cite{2006A&A...459.1007C} to deproject
X--ray imaging and spectral data. It works by interpolating with a
quadratic curve each group of three consecutive points in the profile
and then minimizing the second derivative (i.e. the curvature) of such
curves at the position of the middle point (see A07 for a more
detailed description). The general effect of the regularization is to
smooth out oscillations in the profiles, which are due either to
genuine substructures or to noise which propagates from adjacent bins
in the deprojection.

We define a joint likelihood for the \tsz\ profile,
$\mathcal{L}_{\tsz}$, and for the X--ray surface brightness profile,
$\mathcal{L}_{Xray}$, also including a term associated to the
regularization constraint, $\mathcal{L}_{reg}^\lambda$. Since these
three terms are independent, the total likelihood is given by the
product of the individual ones:
\be
\mathcal{L} \equiv \mathcal{L}_{\tsz} \cdot \mathcal{L}_{Xray} \cdot
\mathcal{L}_{reg}^\lambda. 
\ee 

The fitting parameters are represented by the gas density and total
mass profiles (plus the external pressure). The algorithm first
computes the temperature profile from the density and mass profiles by
inverting the hydrostatic equilibrium equation; then the gas density
and temperature profiles are combined to compute the projected model
profiles of X--ray and \tsz; finally these model profiles are compared
to those obtained from the mock observations to compute the joint
likelihood.  As for the gas density profile, our approach is
completely model--independent, since we treat its value into each
spherical shell as a free parameter, as described in A07. As for the
reconstruction of the total mass profiles, we adopt two different
approaches.  Both methods are described in \cite{2002A&A...391..841E}
for the application to X--ray datasets, although with a different
implementation. Also, \cite{2007MNRAS.380.1521M} present a refined
version of these methods for the study of high resolution X-ray
observations, while \cite{2006MNRAS.369.2013R} discuss their
limitations by applying them to a set of hydrodynamical simulations.

\begin{itemize}
\item {\bf Method 1}. This method consists in the direct
  numerical inversion of the HE equation.  It does not assume any
  particular functional form for the mass profile. Instead, the
  integrated mass enclosed by the mean radius of each shell is treated
  as a free parameter. The only constraint that we impose is that the
  mass has to increase with radius, in order to avoid unphysical
  solutions. The advantage of this method is that it provides a
  completely model--independent reconstruction of the mass profile,
  which relies only on the assumption of hydrostatic equilibrium.
\item{\bf Method 2}. This method is based on assuming the functional
  form for the mass profile provided by the NFW model. Its major
  advantage with respect to Method 1 is that the reconstruction
  becomes more stable, at the cost of assuming a particular model for
  the mass profile. The NFW model is widely adopted in the mass
  reconstruction from X--ray observations. These analyses showed that
  it provides a remarkably good description of the mass profiles of
  massive galaxy clusters, out to large portions of their virial radii
  \citep[see also, e.g.][]{2005A&A...435....1P, 2006ApJ...640..691V,
  2007ApJ...669..158G}. The NFW expression for the mass profile reads
\be
M (<r) = 4\pi r_s^3 \rho_{crit} \delta_c f(x) 
\ee 
where $r_s$ is a characteristic scale length, $x=r/r_s$ is the
distance from the halo centre in units of $r_s$, $f(x)= ln(1+x)
-x/(1+x)$, $\rho_{crit}$ is the critical cosmic density and $\delta_c$
is a characteristic overdensity. It is common (and more convenient) to
rewrite the above equation by expressing $\delta_c$ as a function of
the concentration parameter $c=r_{\Delta}/r_s$, where $\Delta$ is a
given overdensity. Here and in the following, we adopt for $\Delta$
the value of the density contrast at virialization predicted by the
spherical collapse model, which corresponds to $\Delta\simeq 100$ in
our cosmology \citep[e.g.,][]{1996MNRAS.282..263E}. The characteristic
overdensity is then written as
\be 
\delta_c = {\Delta \over 3} {c^3 \over f(x = c)} \,,
\ee
so that the NFW profile becomes
\be
M(<r)={4\pi \over 3} (r_s c)^3 \rho_{crit}\Delta {f(x) \over f(x=c)} \equiv M_\Delta {f(x) \over f(x=c)} \,.
\ee
\end{itemize}

\vspace{0.3truecm}
As for the temperature profile, it is then computed in both methods
from the model mass profile, under the HE assumption (see
Section \ref{sec:hyd-equil}). By inverting the HE equation one first
obtains the gas pressure profile and then the gas temperature, by
combining pressure with gas density. In this way we obtain
\be
kT(r)=- {1\over n_e(r)}\left\{ G \mu m_p \int_{r_{out}}^r {n_e 
M(<r)
\over r^2} dr - P(r_{out})\right\}\,.
\ee
In the above equation the integral is performed from the outermost
radius from which the deprojection is performed, $r_{out}$, and the
radius of interest $r$, while $P_{out}$ is the electron pressure at
$r=r_{out}$. In both methods, we apply the regularization constraint
to the temperature profile, in order to smooth out spurious
fluctuations which are essentially due to the presence of noise.  The
regularization term in the likelihood is given by (see A07 for further
details):
\be
ln(\mathcal{L}_{reg}^\lambda) \equiv - \lambda \sum_{i=3}^{N-1} \left({ 2f_i
-f_{i-1} -f_{i+1}}\right)^2
\label{eq:like_reg}
\ee
where $\lambda=2.5$ and $f_i$ is the value of the temperature in keV
in the $i$-th bin.

Since the hydrostatic equilibrium equation constraints only the
pressure difference between two points, it is necessary to introduce a
further parameter $P(r_{out})$ \citep[see
  also][]{2007MNRAS.tmp..541M}. In particular, in the case of Method 1
this parameter is completely degenerate with the mass enclosed in the
outermost bin. The mass has only a lower boundary, given by the fact
that it cannot be lower than the mass enclosed by the inner bin. This
turns into an instability of the fit which generates an overestimate
of the global mass. This problem is solved by applying the
regularization constraint to the pressure profile.  In this way, in
Method 1 the regularization constraint is actually applied in two
places. Note that once the temperature profile is regularized, the
pressure profile becomes regular as well, so its contribution to the
global likelihood turns out to be generally small. The regularization
of the pressure profile is more important in the outermost two bins,
where it is used to break the degeneracy between $P_{out}$ and the
total mass. So, in Method 2 we insert in the likelihood a second
regularization term, which has the same structure of \eq{eq:like_reg},
where now $\lambda=5$ and $f_i$ is the logarithm of the pressure in
the $i$-th bin. Note that the numerical value of $\lambda$ does not
give a direct indication of the weight given to regularization,
because this depends on the units used for the quantity $f_i$.

Given the high number of fitting parameters ($2N+1$ for Method 1,
$N+3$ for Method 2 respectively), we apply a MCMC technique, which is
described in A07. The technique has the advantage of computing the
(marginalized) probability distribution of all parameters
simultaneously, together with the whole covariance matrix. For a more
detailed information on MCMC fitting techniques we refer to
\cite{1993Neal}, \cite{gilks1996} and \cite{mackay1996}.

\begin{figure*}
\centerline{
\psfig{file=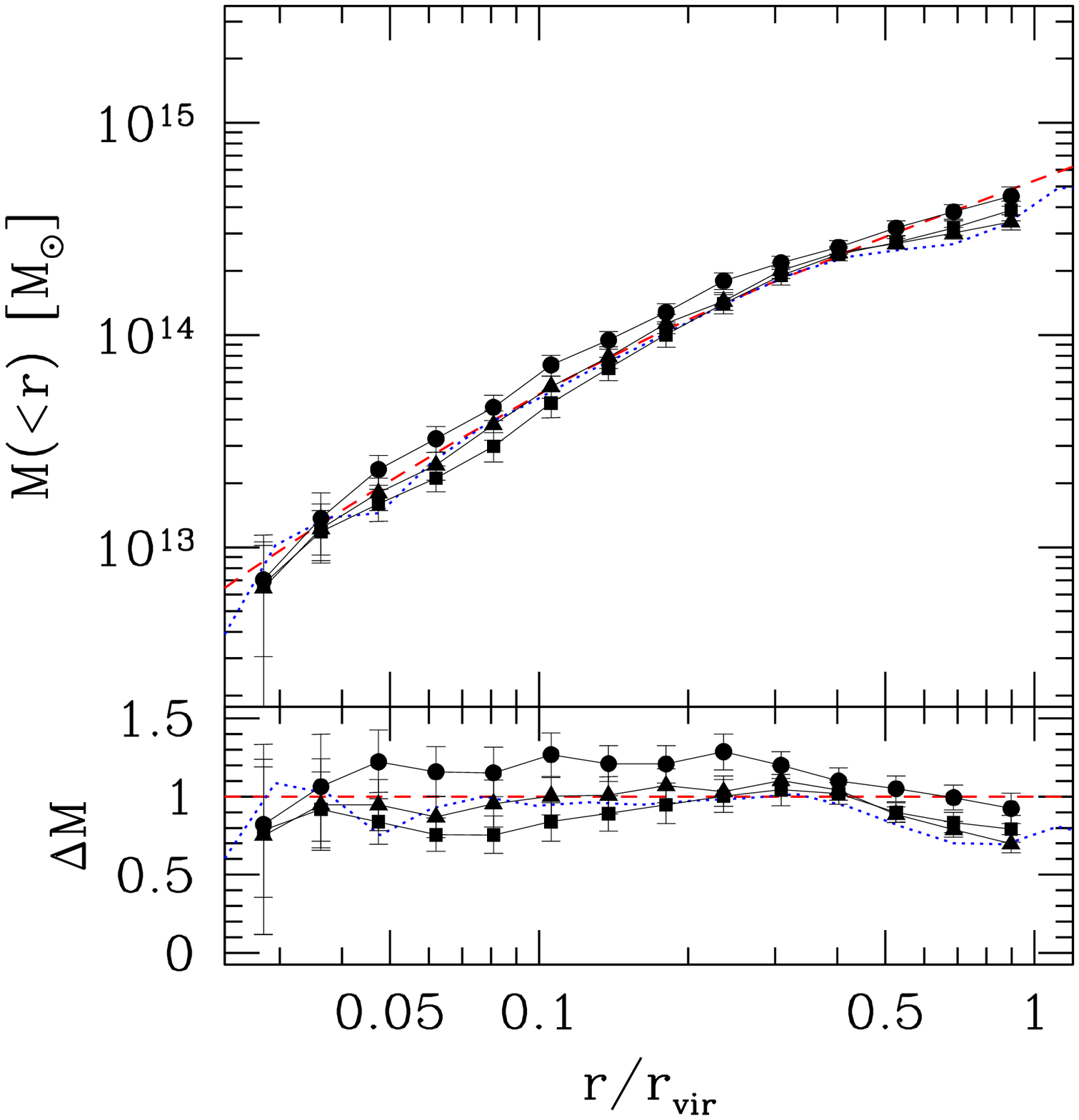,width=\fwid}
\psfig{file=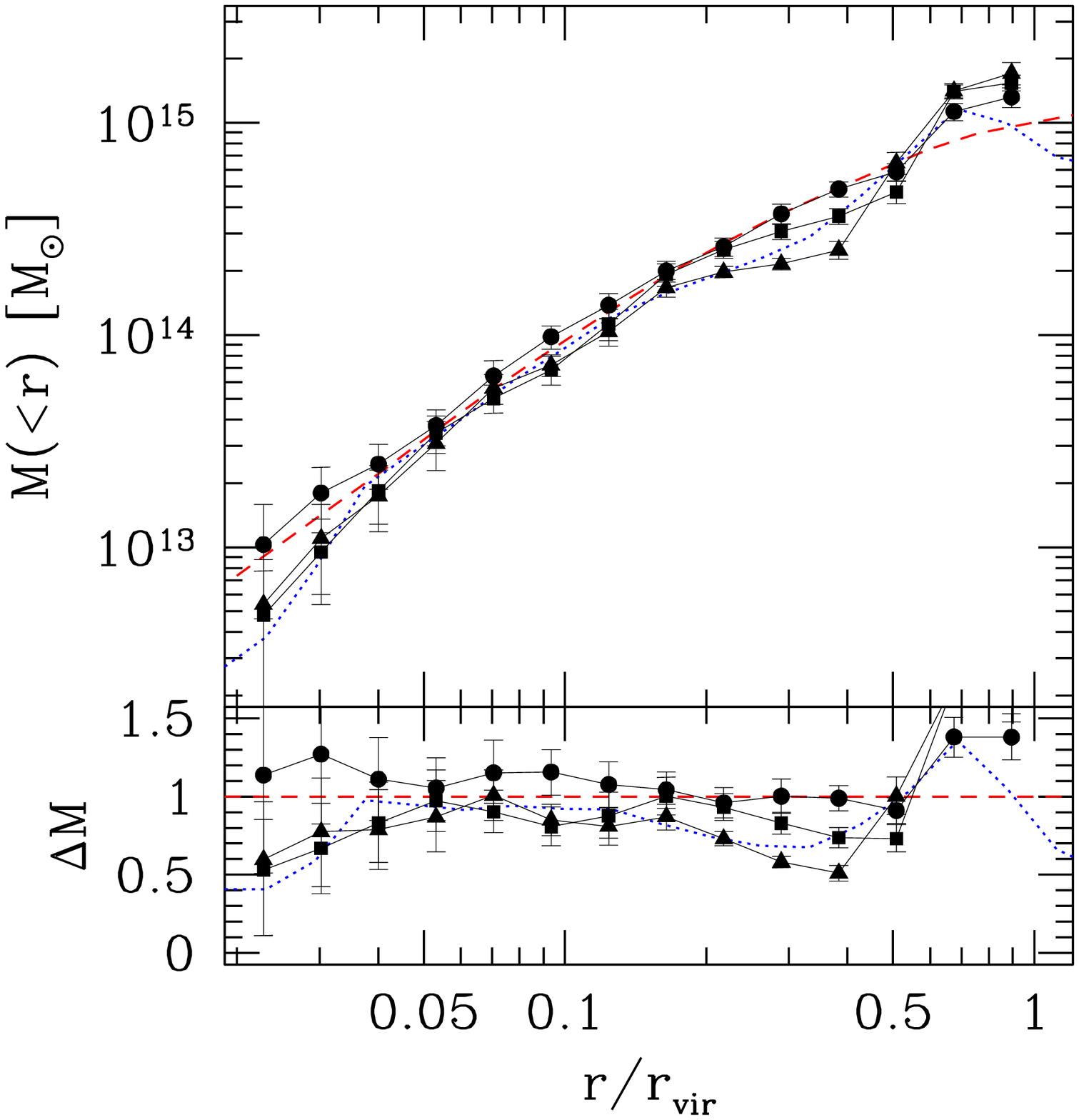,width=\fwid}
}
\centerline{
\psfig{file=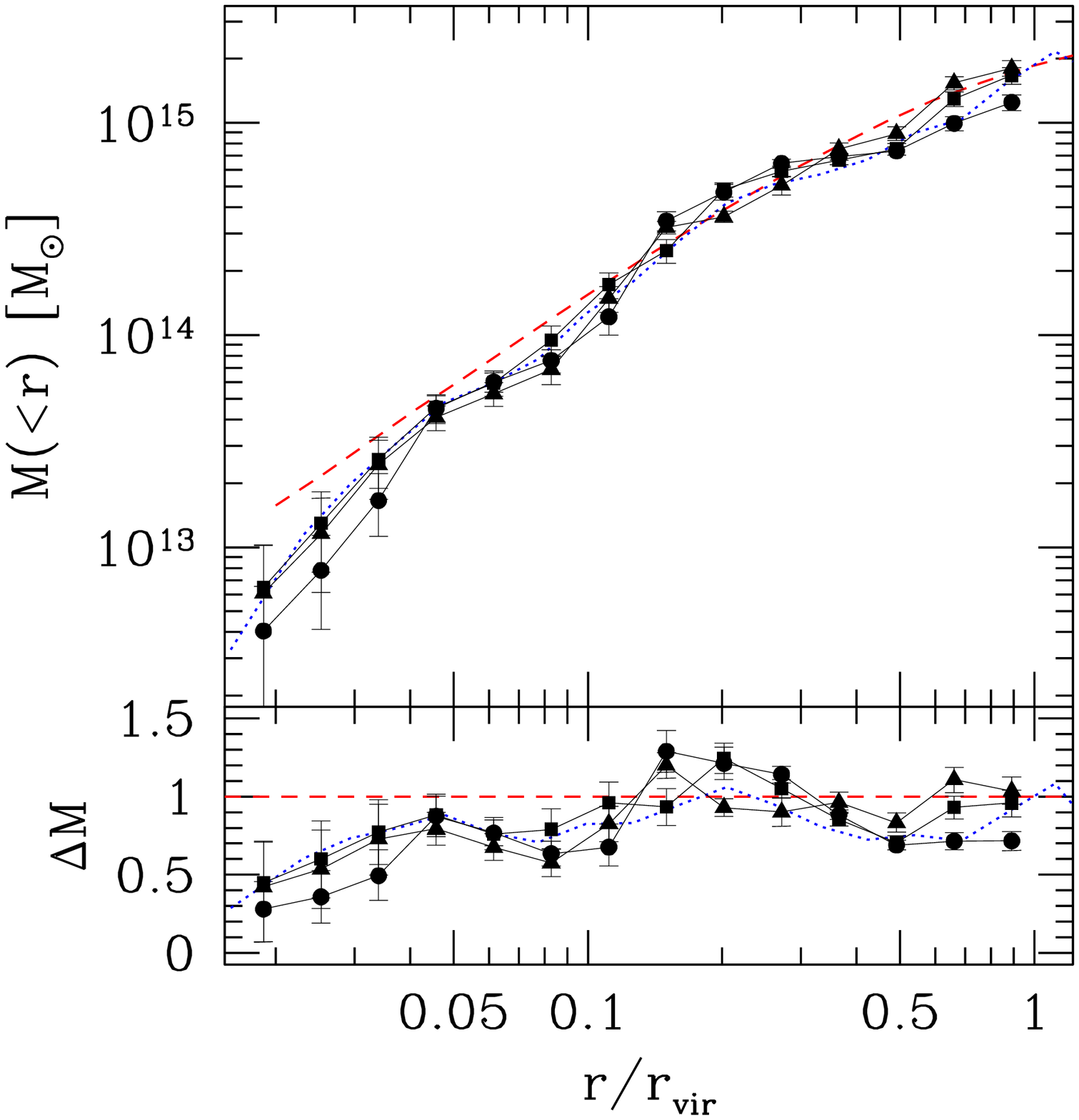,width=\fwid}
\psfig{file=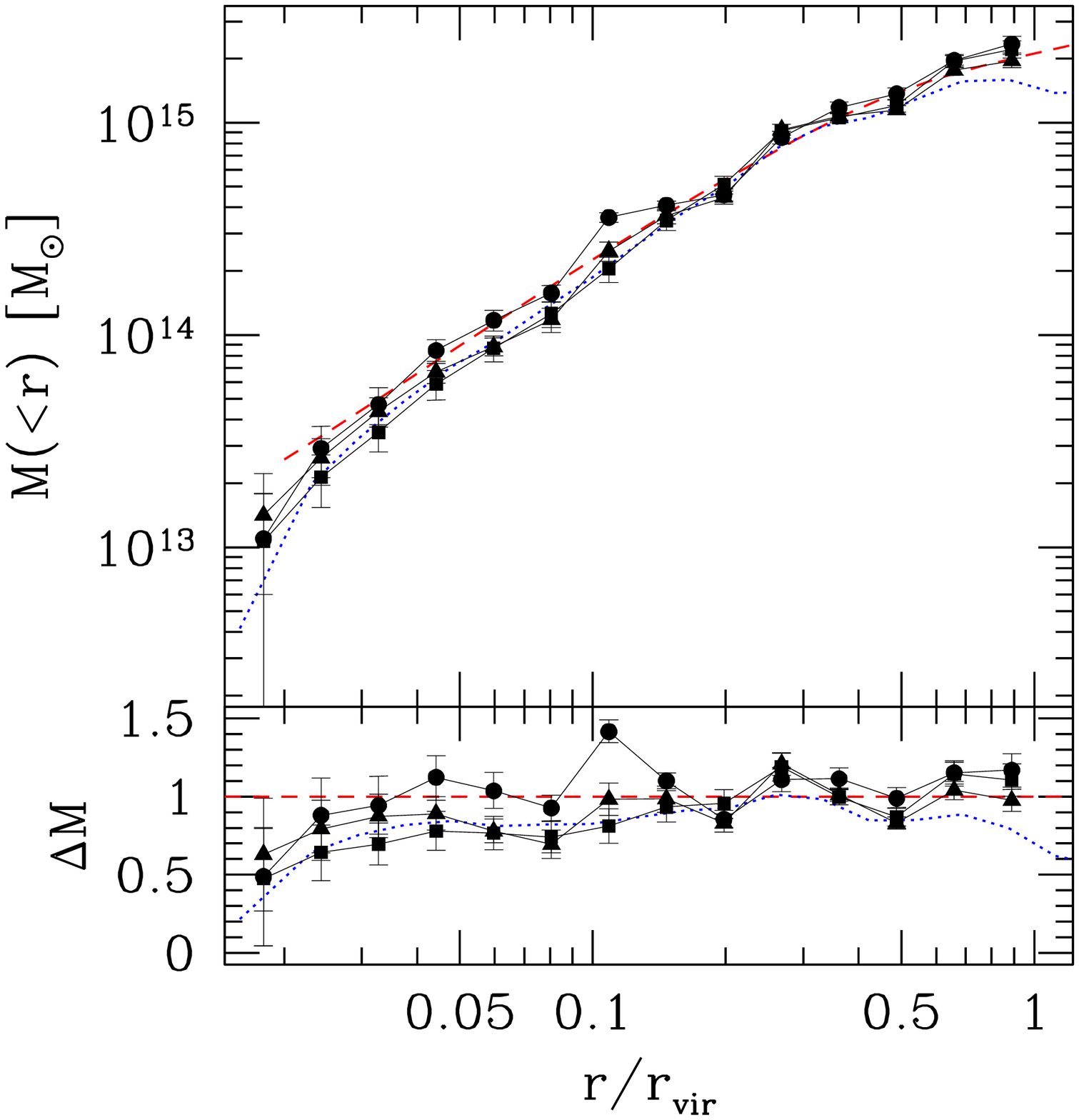,width=\fwid}
}
\caption{Mass reconstruction for the clusters C1 (upper left), C2
  (upper right), C3 (lower left) and C4 (lower right), while applying
  Method 1. The black triangles (squares, circles) and
  line represent the reconstruction along the $x$ ($y$, $z$) axis. The
  red dashed line represents the true mass profile, while the blue
  dotted line is hydrostatic mass obtained from the application of
  \eq{eq:hydeq} to the true 3--dimensional gas density and temperature
  profiles. The lower part of each panel shows the ratio between the
  same quantities and the true mass.}
\label{fi:mass-free}
\end{figure*}

\section{Results} \label{sec:total-results} 
As already pointed out, both methods of mass reconstruction adopt a
different implementation with respect to the deprojection technique
presented in A07.  Reassuringly enough, the gas density and
temperature profiles obtained from all methods are virtually
identical. For this reason, they are not shown in the following while
we will concentrate our discussion only on the mass reconstruction.

\subsection{Total mass profiles: method 1}

We show in \fig{fi:mass-free} the comparison between the reconstructed
and the true mass profiles for the C1--4 clusters, chosen as
examples. We generally find an underestimate by about 10--15 per cent
throughout the virial radius, with slightly larger deviations in the
centre and in the outskirts. For all these four clusters (and for most
of the others) the uncertainties in X--ray and tSZ profiles do not
allow us to place tight constraints on the amount of mass contained
within the innermost bin. Since this quantity has no lower boundary
(e.g. given by the mass contained within an inner bin), we find cases
in which the lower limit of this mass takes very low or even negative
values.  To avoid this problem, we fix by hand a lower limit for the
mass within the innermost bin at $10^{12}\msun$.  A comparison of the
reconstructed mass profiles with the profiles of hydrostatic mass,
$M_{hyd}$, generally show a close agreement.  This demonstrates that
the main source of systematics is intrinsic, namely the deviation of
the gas from a perfect hydrostatic equilibrium, while the deprojection
method is essentially unbiased. While this is strictly true out to
$r\simeq 0.5r_{vir}$, we note a tendency for the reconstructed mass to
lie above $M_{hyd}$ when approaching the virial radius. We attribute
this to the lower signal--to--noise ratio in the external regions
(SNR~$\sim 4-5$ for both the SZ and X--ray signals in the outermost
bin, compared to 15--20 in the innermost bins) and to a larger impact
of fore/background contaminations in these regions. In fact, the
recovered mass at these radii is bound not to be smaller than the mass
at the inner radii, while it does not have any upper boundary.  As a
consequence, any deviation associated to noise or contamination can
only act in the direction of increasing the mass.  We further point
out that our synthetic X--ray maps do not include a background, so the
noise here is only given by the poissonian noise of the
signal. Clearly, the inclusion of a realistic background would limit
our analysis to scales comparable to those reached by X--ray
observations, which only in some cases trace the emissivity beyond
$\sim r_{200}$ \citep[][]{2005A&A...439..465N}.

\begin{figure}
\centerline{
\psfig{file=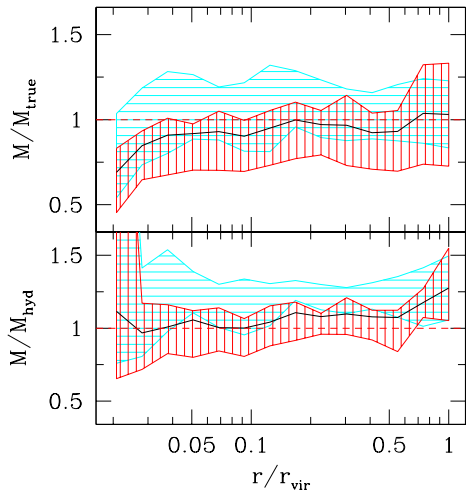,width=\fwid}
}
\caption{Reconstructed vs. true (upper panel) and hydrostatic (lower
  panel) mass, averaged over the set of simulated clusters, while
  applying Method 1. In each panel, the horizontally (vertically)
  shaded area represents the mean$\pm 1 \sigma$ over the projections
  along the $z$ ($x$ and $y$) axis. The black line represents the mean
  over all the projections of all clusters.}
\label{fi:convol-free}
\end{figure}

These results are also confirmed from the analysis of the whole set of
simulated clusters. Indeed, \fig{fi:convol-free} shows that in the
inner regions (out to about $r=0.15 r_{vir}$) the mass profile
recovered from our deprojection algorithm are quite close on average
to the hydrostatic masses, while being consistently smaller that the
true mass profiles, by about 10--15 per cent. Instead, in the
outskirts the reconstructed mass profiles seem to well recover the
true ones. However, this agreement is due to the effect of two biases,
namely a violation of HE and the effect of noise in the deprojection
algorithm, which act in opposite direction, rather than to an
intrinsically better performance of the deprojection method at large
radii.

In A07 we showed that the C1 cluster has a significant contamination
in the projection along the $z$ axis, due to a merging group which is
approaching the cluster virial region. The effect of this
contamination was to boost the temperature profile by about 20 per
cent. In the upper left panel of  \fig{fi:mass-free} we show
that this causes a mass overestimate by a comparable amount of the
mass reconstructed from the projection along the $z$ axis, which lies
above the hydrostatic mass in this particular case. 
Also for the C2, C3 and C4 clusters the mass reconstructed from the
projection along the axis of maximum elongation is larger than the
mass recovered from the other two projections. \fig{fi:convol-free}
confirms that this behaviour is still present when averaging
over the whole set of simulated clusters. 

\subsection{Total mass profiles: method 2} \label{sub:results:met2}

\begin{figure*}
\centerline{
\psfig{file=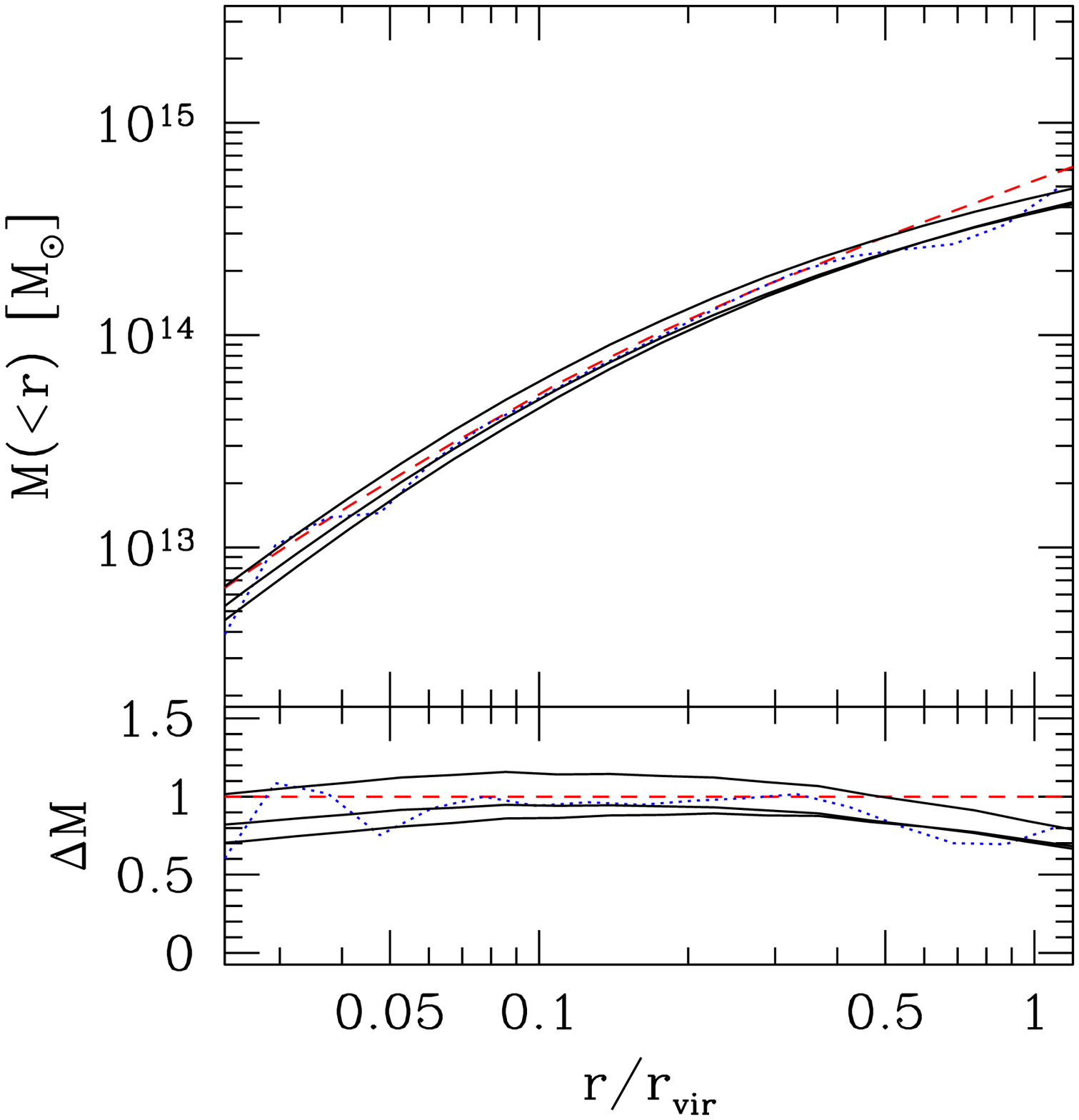,width=\fwid}
\psfig{file=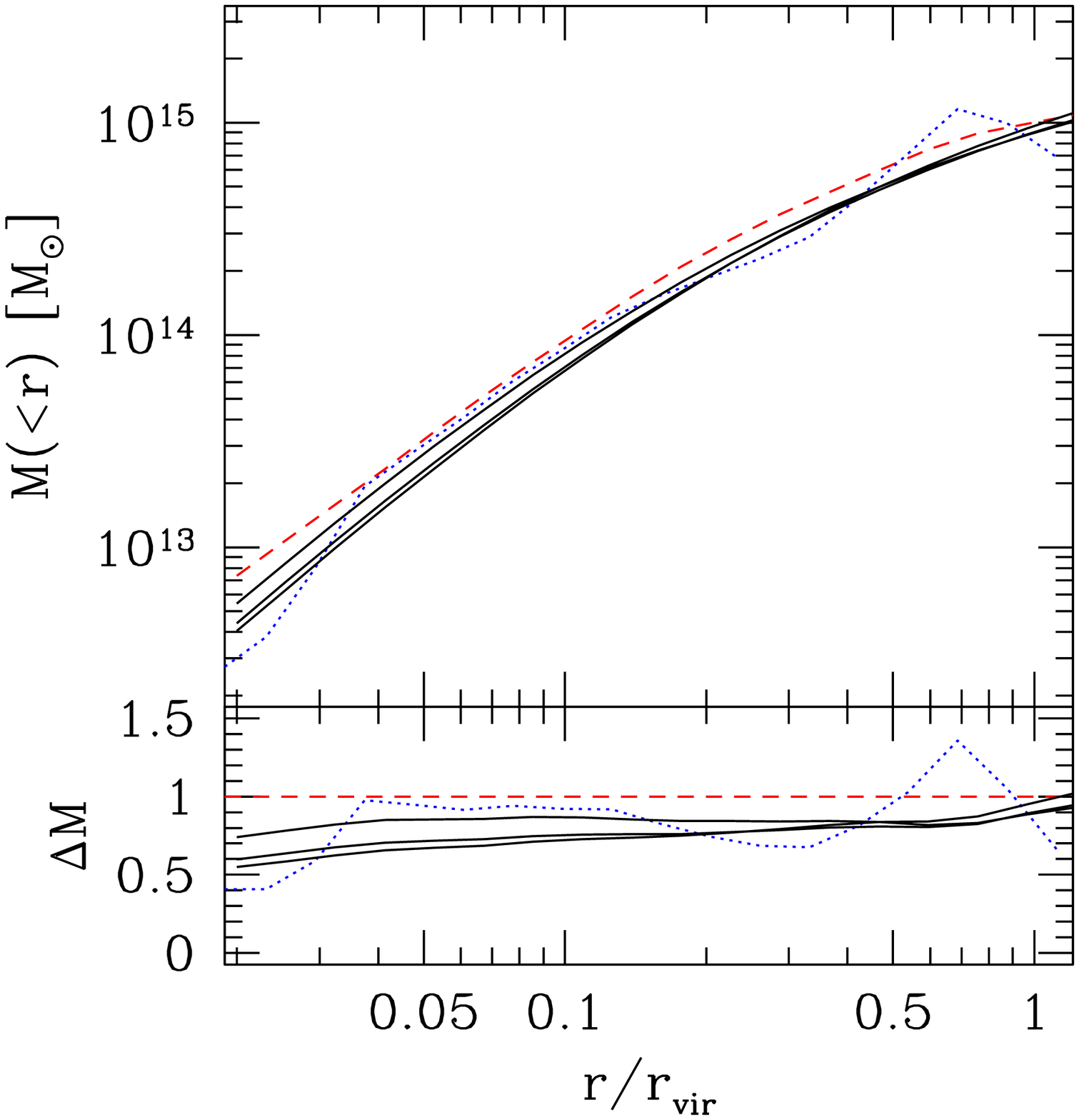,width=\fwid}
}
\centerline{
\psfig{file=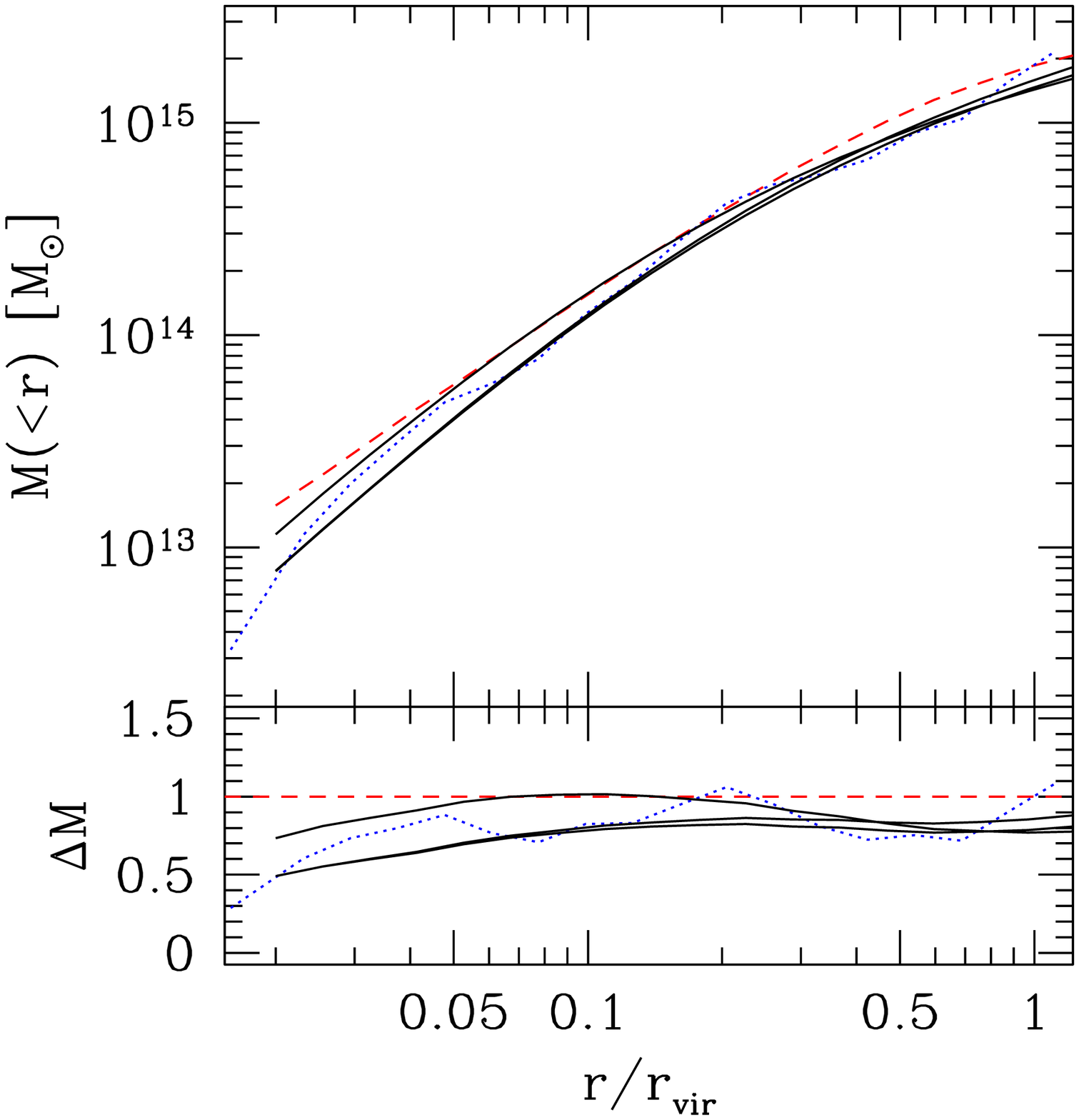,width=\fwid}
\psfig{file=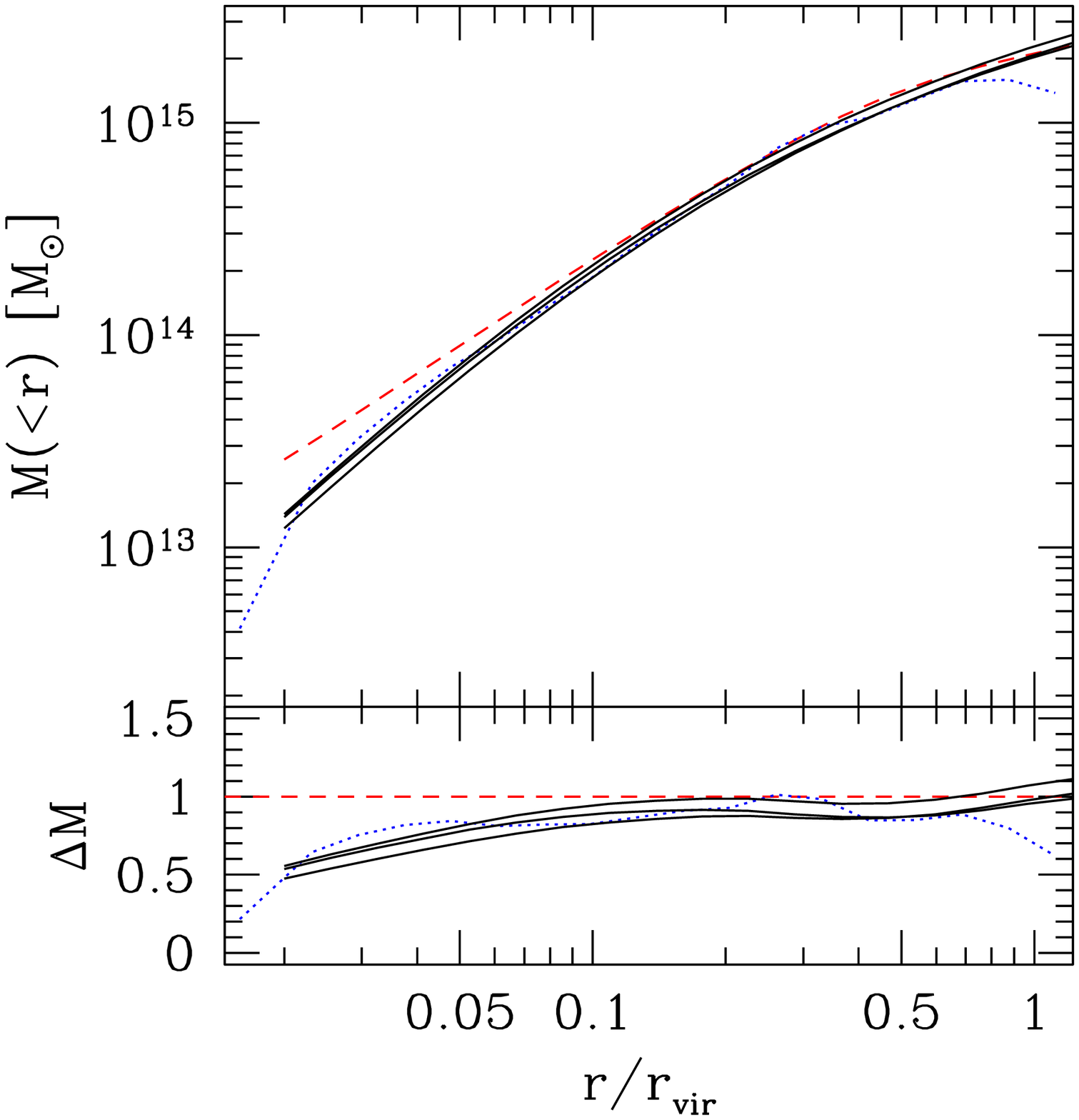,width=\fwid}
}
\caption{The reconstruction of the mass profiles for the C1, C2, C3
  and C4 clusters, based on the Method 2, compared with the true
  profiles (dashed curves) and with the profiles recovered from the
  hydrostatic equilibrium (dotted curves).  For each cluster, the
  three solid curves show the reconstruction along the three axes of
  projection.} \label{fi:mass-nfw}
\end{figure*}

We show in \fig{fi:mass-nfw} the mass profiles of the same 4
clusters chosen as examples, as reconstructed from the application of
the method based on assuming an NFW model. In this case, the
reconstructed profiles are not reported with errorbars, which are
assigned to the total mass $M$ and to the concentration parameter $c$,
which are the only two parameters entering in the fitting
procedure. The accuracy in recovering these parameters is discussed in
the following.
\begin{table}
\centerline{
\begin{tabular}{llcccccc}
&$\Delta$& $\displaystyle{ M_{\Delta}(<r_{\Delta,sim}) \over
    M_{\Delta,true}}$ & $\displaystyle{r_{\Delta,fit}\over
    r_{\Delta,true}}$ & $\displaystyle{M_\Delta(r_{\Delta,fit})\over
    M_{\Delta,true}}$ \\ \hline \hline $r<r_{vir} $&vir & $0.81 \pm
  0.10 $ & $0.91 \pm 0.05 $ & $0.77 \pm 0.11 $\\ \hline
  $r<r_{500}$&500& $0.89 \pm 0.16 $ & $0.95 \pm 0.08 $ & $0.87 \pm
  0.20 $ \\ &vir& $0.90 \pm 0.21 $ & $0.95 \pm 0.10 $ & $0.89 \pm 0.27
  $\\
\end{tabular}
}
\caption{Summary of the results of Method 2 over the
whole set of simulated clusters. The first column reports the limiting
radius of the fit, while the second column reports the overdensity $\Delta$
at which all quantities are computed. Columns 3 to 5 report the mean
and standard deviation for the
accuracy in recovering the following quantities: the total mass
within $R_\Delta$ when it is computed from simulation data; the
value of $R_\Delta$ obtained from the fit; the total mass enclosed by
$R_\Delta$ when it is fitted from data (note that in this case the
fitted mass and the true mass refer to different radii). All these
data are computed by excluding the cluster labeled C5 
(see text for details).}
\label{tab:res}
\end{table}

\begin{figure*}
\centerline{
\psfig{file=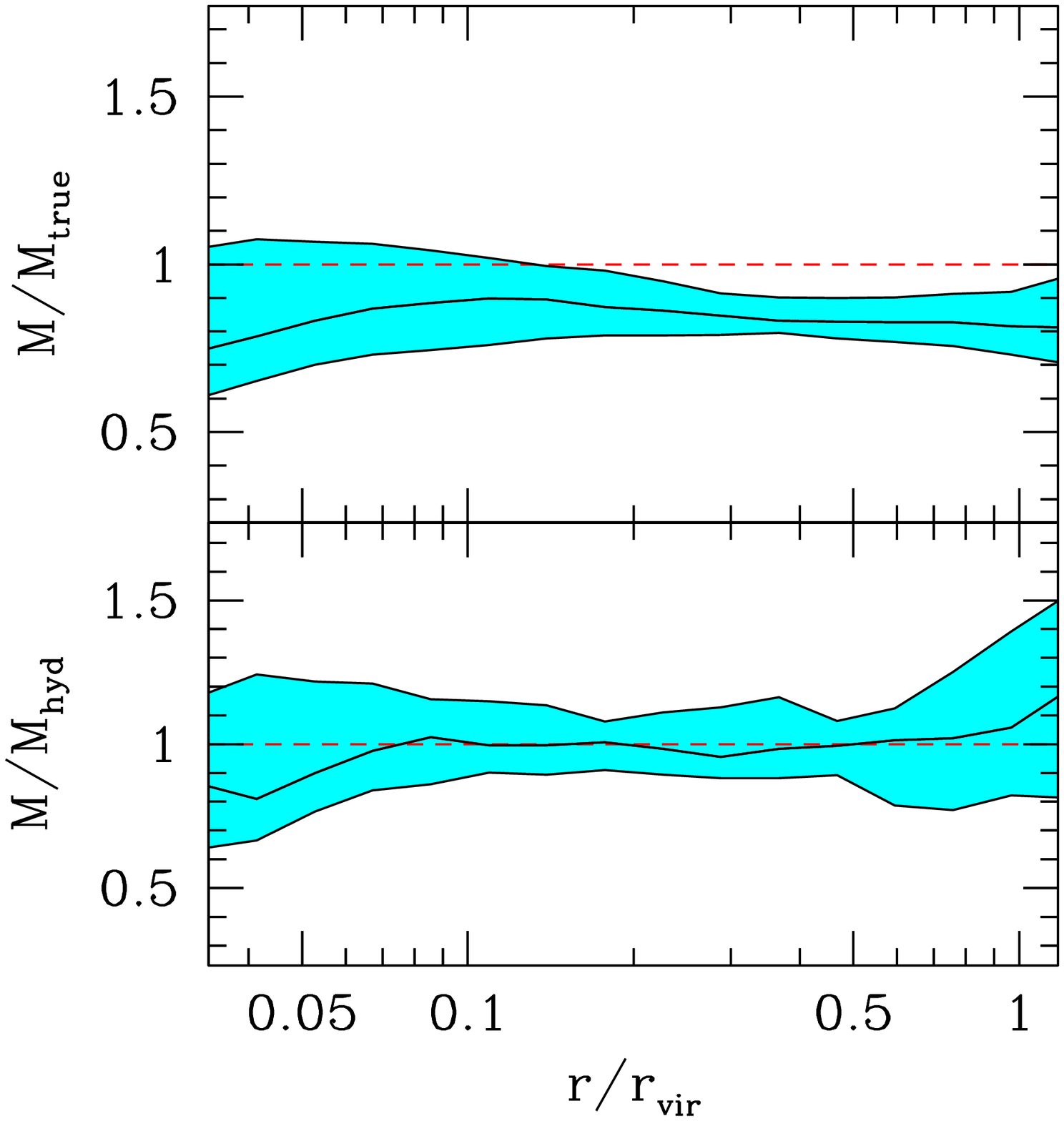,width=\fwid}
\psfig{file=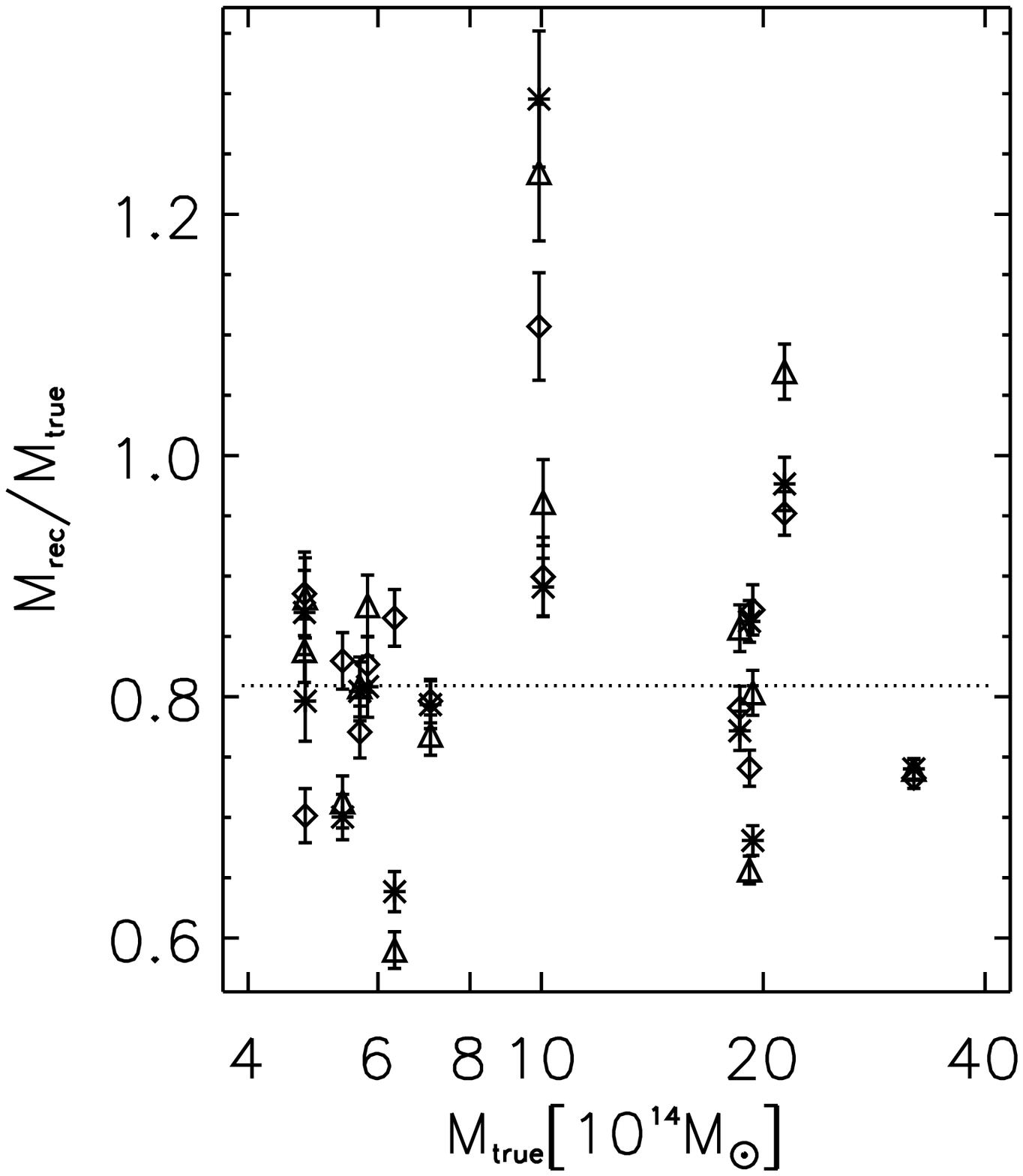,width=\fwid}
}
\caption{Left panel: the ratio of the reconstructed mass profiles,
  based on Method 2 (fitting out to $r_{vir}$), with the true (upper
  panel) and hydrostatic (lower panel) mass profiles, after averaging
  over the set of simulated clusters.  The shaded area encompasses 68
  per cent of the recovered profiles, while the solid line shows the
  median profile. Right panel: the ratio between the recovered and the
  true masses, both computed within $r_{vir}$, as a function of the
  true mass of each cluster. Diamonds, triangles and stars are for the
  reconstruction from the projection along the $x$, $y$ and $z$ axis
  respectively. Errorbars represent the 1$\sigma$ statistical
  uncertainty on the recovered mass, due to the noise in mock
  SZ/X--ray images. The horizontal dotted line indicates the mean
  deviation of the reconstructed mass.}
\label{fi:convol-nfw-rvir}
\end{figure*}

Clearly, assuming an analytical form for the mass profiles has the
advantage of providing a much more stable reconstruction and a smooth
profile. Similarly to the case of the Method 1, the recovered mass is
much closer to the hydrostatic mass than to the true one.  The left
panel of \fig{fi:convol-nfw-rvir} shows the accuracy of the recovered
mass profile, after averaging over the whole set of simulated
clusters. From the upper panel of this plot, we note that the mass
profile is generally underestimated by about 15--20 per cent. The
slightly larger underestimate with respect to Method 1 is due to the
fact that here we are adopting an analytical model. We remind that the
violation of hydrostatic equilibrium is larger at $r_{vir}$, which is
the outermost radius from which the reconstruction of the mass profile
begins. Since we force the profile to follow the NFW shape when
reconstructing at smaller radii, this larger mass underestimate is now
propagated inward. This is different from what happens for the Method
1, where the value of the mass at an inner radius is not forced by the
extrapolation of the profile recovered at a larger radius.  The right
panel of \fig{fi:convol-nfw-rvir} shows the accuracy in recovering the
total mass as a function of the true mass of the cluster. We find that
the statistical uncertainties in the estimate of the cluster masses
are generally smaller than the bias induced by the violation of the
hydrostatic equilibrium and are also smaller than the scatter
associated to the choice of the projection direction. This panel also
shows the presence of an outlier, which is the cluster labeled C5;
this object has a much more irregular structure when compared with the
rest of the sample. For this reason, here and in the following we
exclude it from the computation of the mean and the standard deviation
over the sample of simulated clusters, as reported in
\tab{tab:res}. Within $r_{vir}$, we find $M_{rec}/M_{true}= 0.81 \pm
0.10$, with no significant dependence on the cluster mass.

\begin{figure*}
\centerline{
\psfig{file=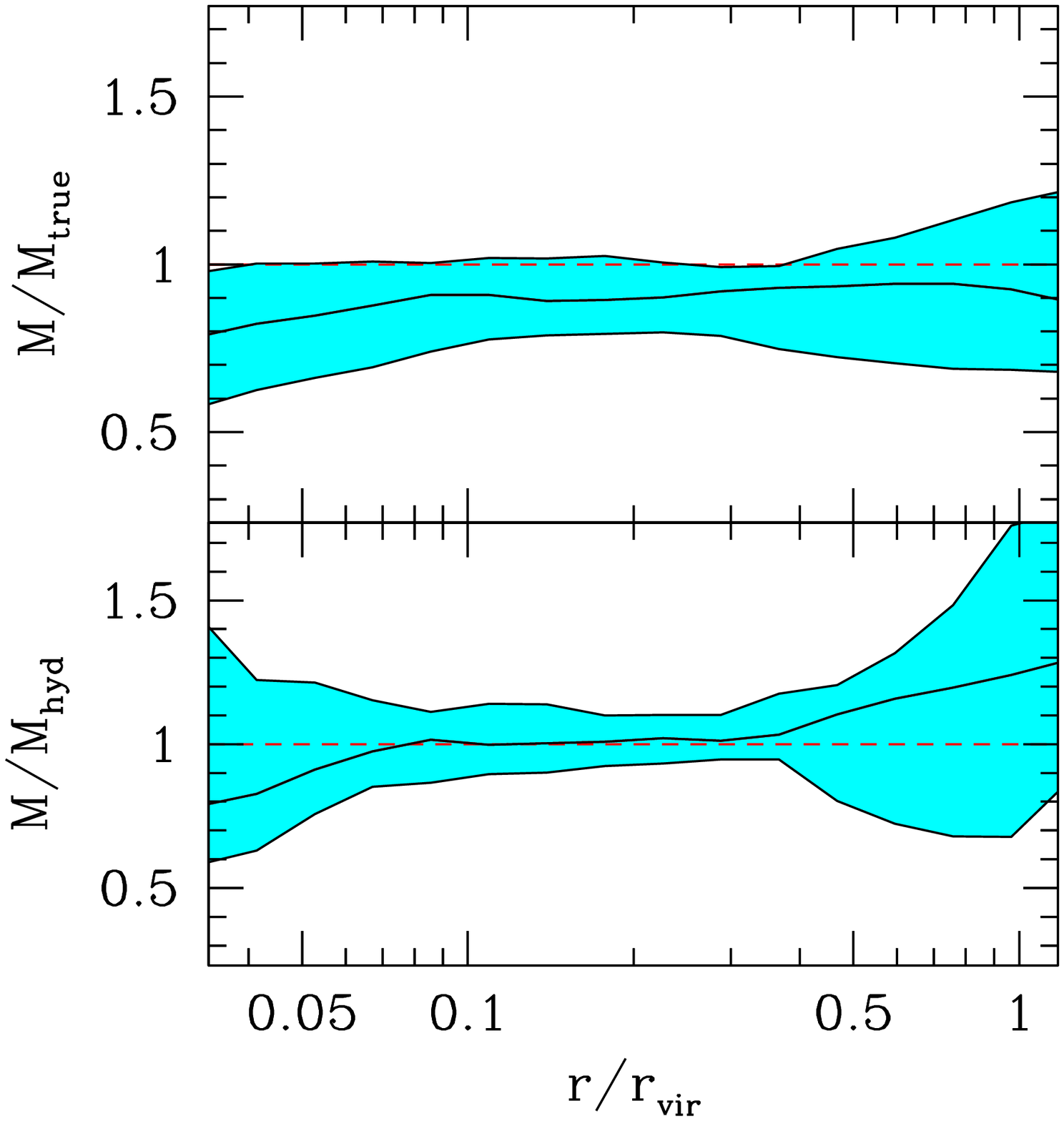,width=\fwid}
\psfig{file=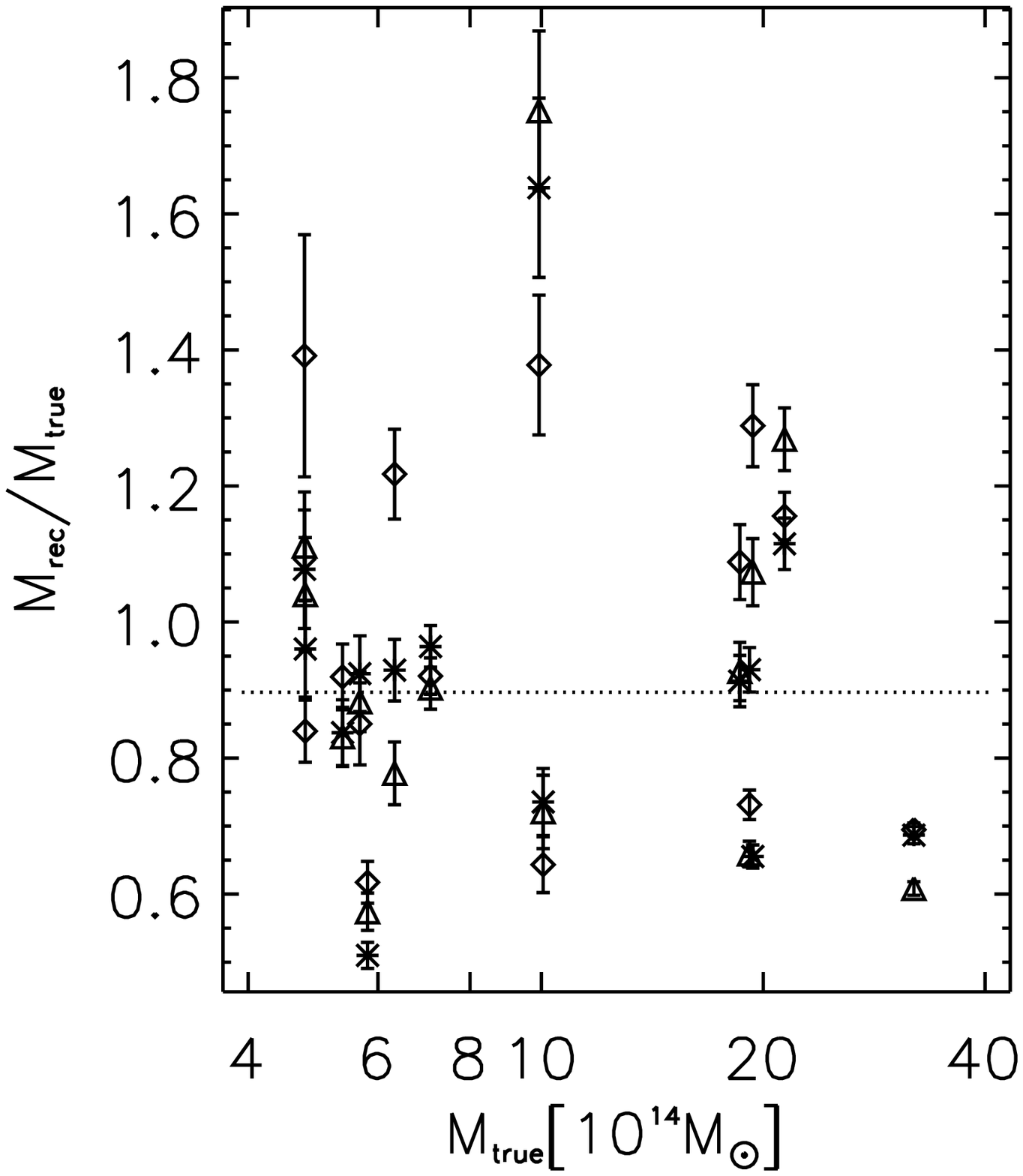,width=\fwid}
}
\caption{The same as in \fig{fi:convol-nfw-rvir}, but fitting the NFW
profile out to $r_{500}$, and extrapolating it out to $r_{vir}$.  }
\label{fi:convol-nfw-r500}
\end{figure*}

The increase of the HE violation at large radii suggests that the mass
profiles should be better recovered in normalization by limiting the
fit to a smaller region, typically $\mincir r_{500}$. This radius
represents the outermost limit typically reached by X--ray
observations. We repeat the analysis by limiting the fit to $r_{500}$,
while using the recovered NFW model to extrapolate the mass profile
out to the virial radius. The left panel of \fig{fi:convol-nfw-r500}
demonstrates that restricting the reconstruction of the profile within
$r_{500}$ has the expected effect of reducing the mass
underestimate. While the recovered mass profiles are very close to the
hydrostatic ones out to the fitting radius $r_{500}$, it exceeds the
HE predictions at larger radii, while remaining within the 10 per cent
underestimate of the true mass out to the virial radius. This
demonstrates that the NFW model reconstructed within $r_{500}$ can be
safely extrapolated to larger radii to extend the mass profile
reconstruction. After averaging over the set of simulated clusters, we
find that $M_{500}$ and $M_{vir}$ are both recovered with an
underestimate of only 10 per cent (see \tab{tab:res}).  The right
panel of \fig{fi:convol-nfw-r500} shows the accuracy in recovering the
virial mass of each cluster. This plot has a larger scatter with
respect to \fig{fi:convol-nfw-rvir}, due to the fact the we are now
fitting over mass profiles over a narrower radial range (see
\tab{tab:res}). In agreement with the results of
\cite{2008arXiv0808.1111P}, we conclude that a reliable procedure is
to reconstruct the mass profiles within relatively small radii, $\sim
r_{500}$, where the HE is not seriously violated, while extrapolating
to so-obtained mass profile according to the NFW model.
\begin{figure*}
\centerline{  
\psfig{file=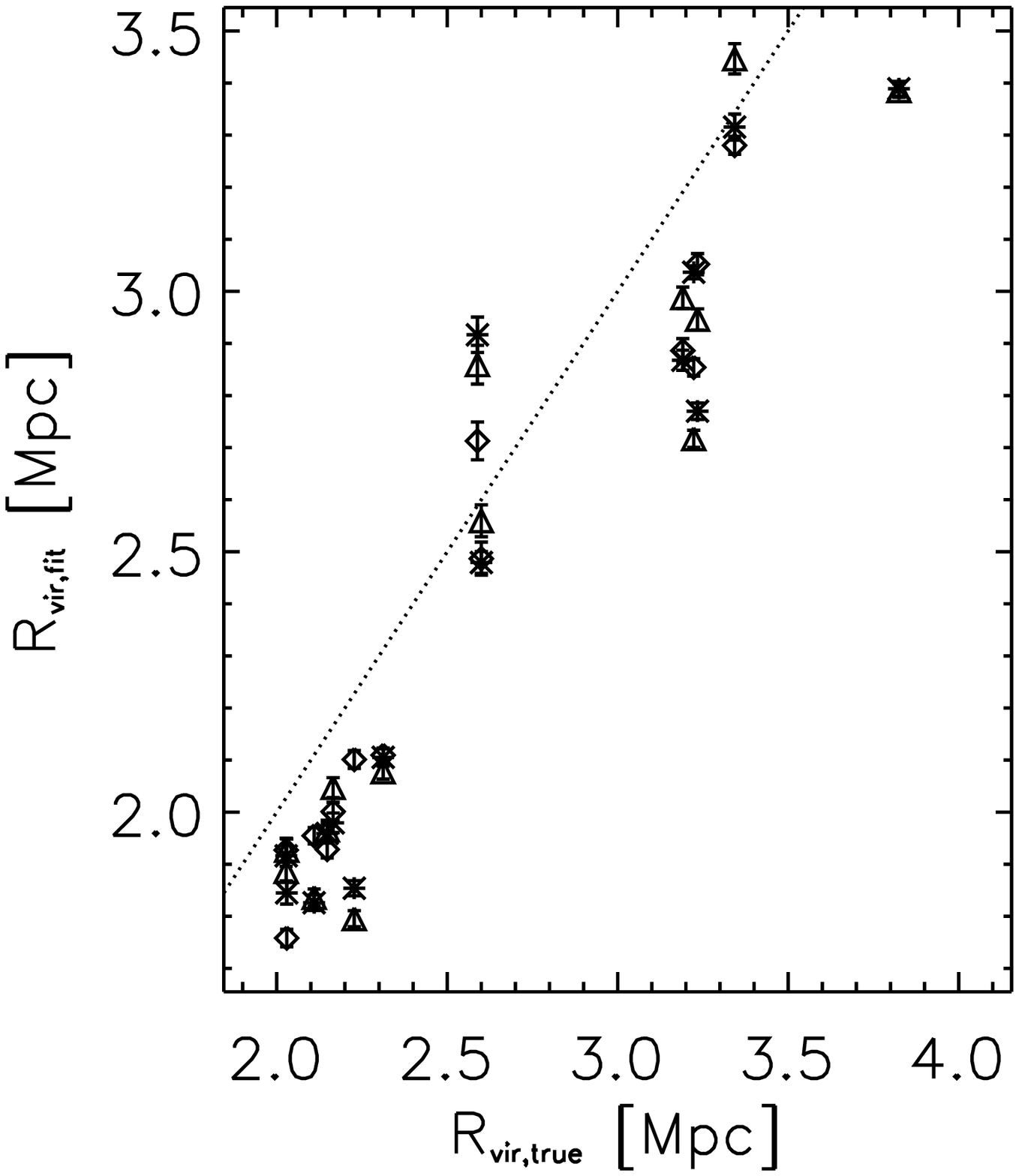,width=\fwid} 
\psfig{file=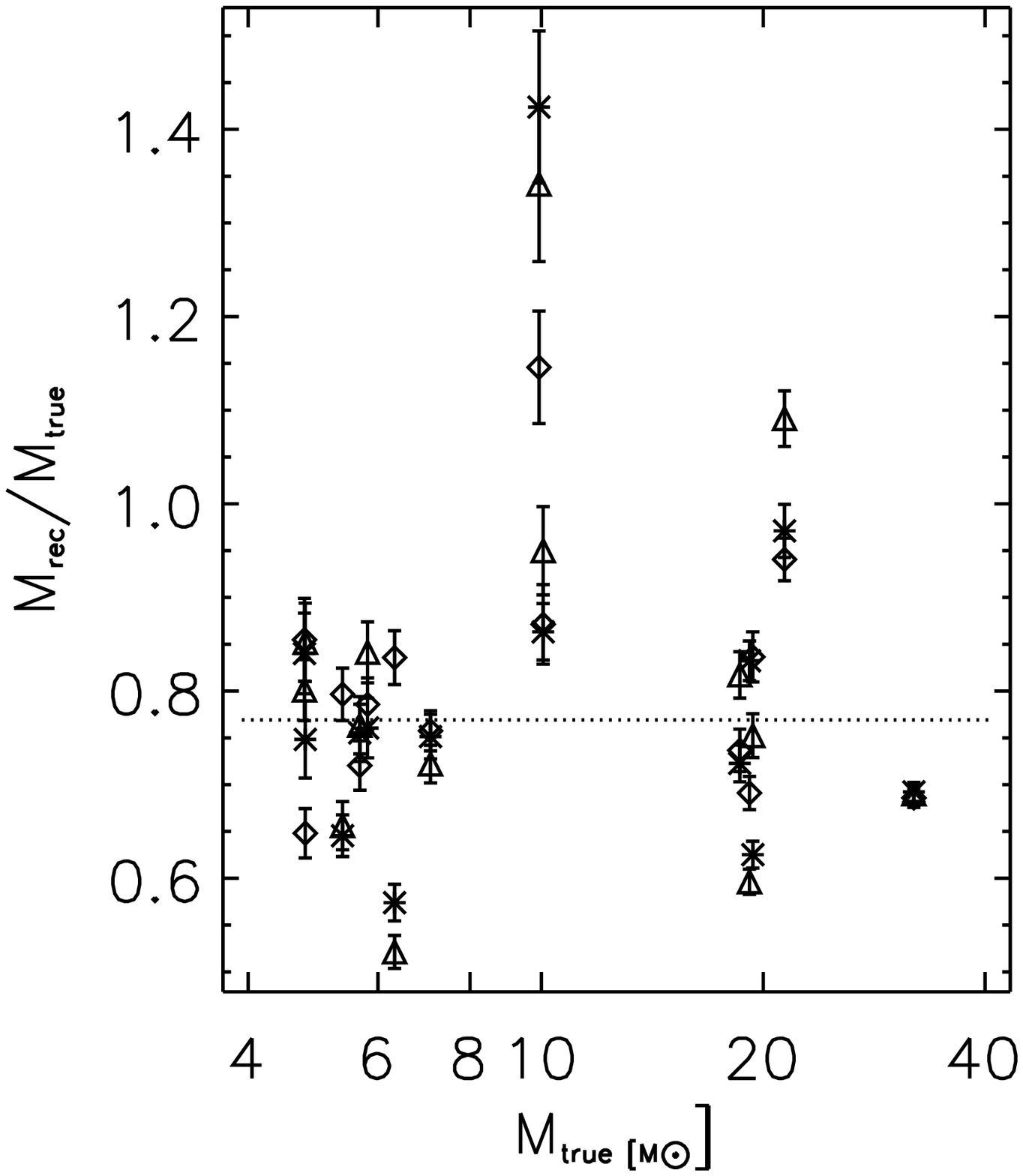,width=\fwid} 
}
\caption{Left panel: the estimated virial radius $r_{vir,fit}$ using
  Method 2 out to $r_{vir}$) vs.  the true one computed from the
  simulation data $r_{vir,sim}$. The dotted line reports the
  one--to--one relation. Right panel: the recovered mass within
  $r_{vir,fit}$ vs. the true mass within $r_{vir,sim}$. In both
  panels, diamonds, triangles and stars are for the reconstruction
  from the projection along the $x$, $y$ and $z$ axis respectively.
  Errorbars represent the 1$\sigma$ statistical uncertainty on the
  recovered mass, due to the noise in mock SZ/X--ray images. The
  horizontal dotted line represents the mean over all the mass
  determinations.}
\label{fi:total-rvir}
\end{figure*}

In the results shown so far, we assumed that we exactly know the value
of the radius, either $r_{vir}$ or $r_{500}$, within which the masses
are recovered.  In the analysis of observational data, instead, the
virial radius is generally not known in advance, but it is estimated
directly from the recovered mass profile. Therefore, if the mass
profile is under/overestimated, also the estimate of the virial radius
will be biased low/high. As pointed out by \cite{2007ApJ...655...98N},
this turns out into a larger bias in the recovered $M_{500}$ or
$M_{vir}$. In order to quantify this effect, we show in the left panel
of \fig{fi:total-rvir} the virial radii inferred from the
reconstructed mass profiles, $r_{vir,fit}$, vs. the true ones
$r_{vir,sim}$. Clearly, the recovered virial radii are generally
smaller than the true one, by about 10 per cent.  This bias decreases
by a few percent if $r_{vir,fit}$ is recovered from the extrapolation
of the profile fitted within $r_{500}$ (see\tab{tab:res}).  The right
panel of \fig{fi:total-rvir} shows the recovered virial mass when
using $r_{vir,fit}$ versus the true virial mass. A comparison of this
plot with the right panel of \fig{fi:convol-nfw-rvir} shows that the
effect of using the recovered virial radius is that of slightly
increasing the underestimate of the virial mass, which is now
$M_{vir,fit}/M_{vir,true}=0.76 \pm 0.11$. Again, we verified that this
effect is reduced, with $M_{vir,fit}/M_{vir,true}=0.89 \pm 0.27$ when
using instead the virial radius extrapolated from the analysis
performed within $r_{500}$.

Finally, we find that observing the same cluster along different lines
of sight cause differences in the reconstructed mass by only a few per
cent (see
\figss{fi:convol-nfw-rvir}{fi:convol-nfw-r500}{fi:total-rvir}).  It is
worth noting that the axes of projection are not chosen randomly,
instead they are fixed to the principal axes of the inertia tensor.
This choice maximizes the difference between the projected images, so
our estimate can be considered an upper limit of this effect.

\section{The concentration--mass relation}\label{sec:cm-rel} 
\begin{figure*}
\centerline{
\psfig{file=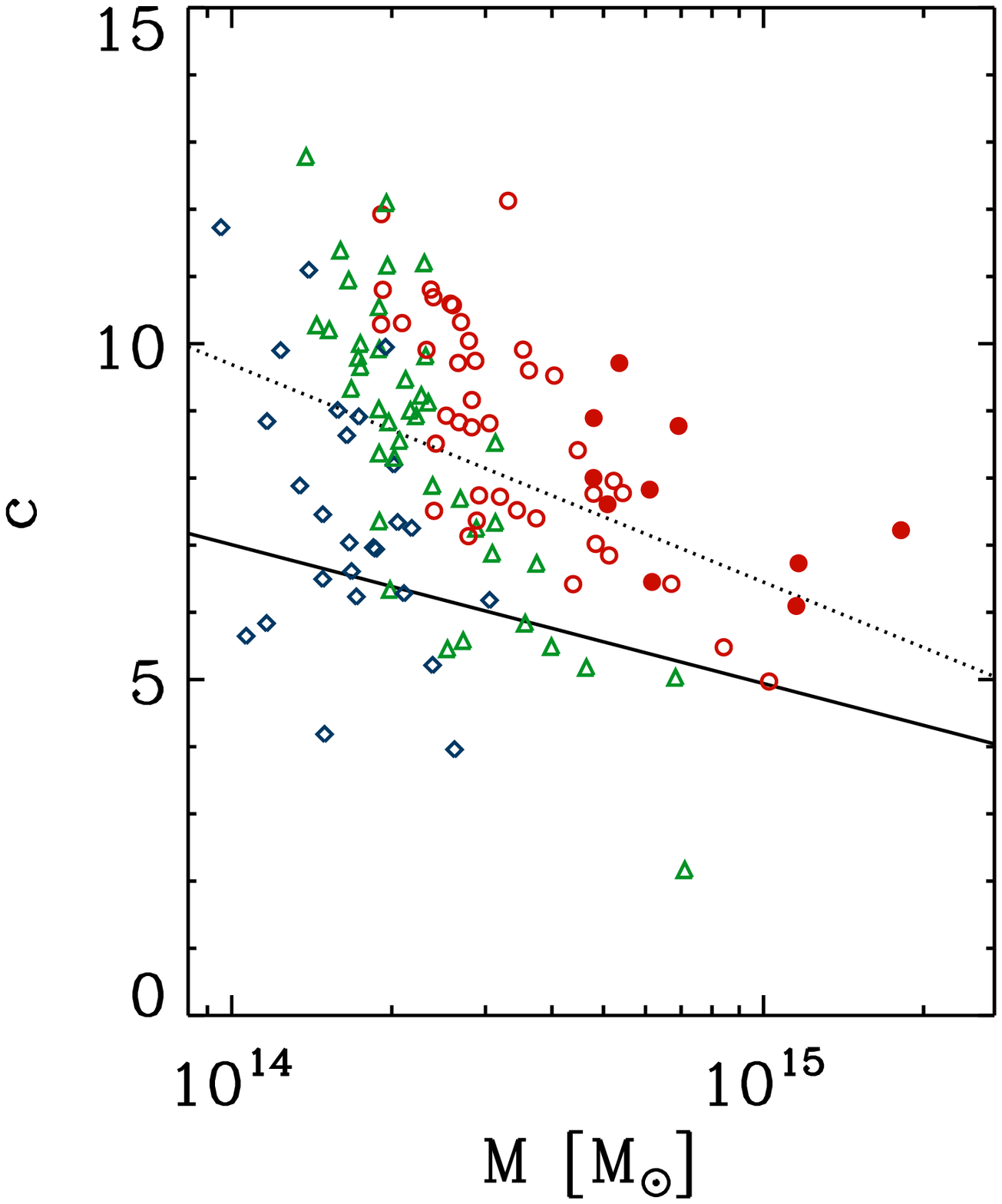,width=\fwid}
\psfig{file=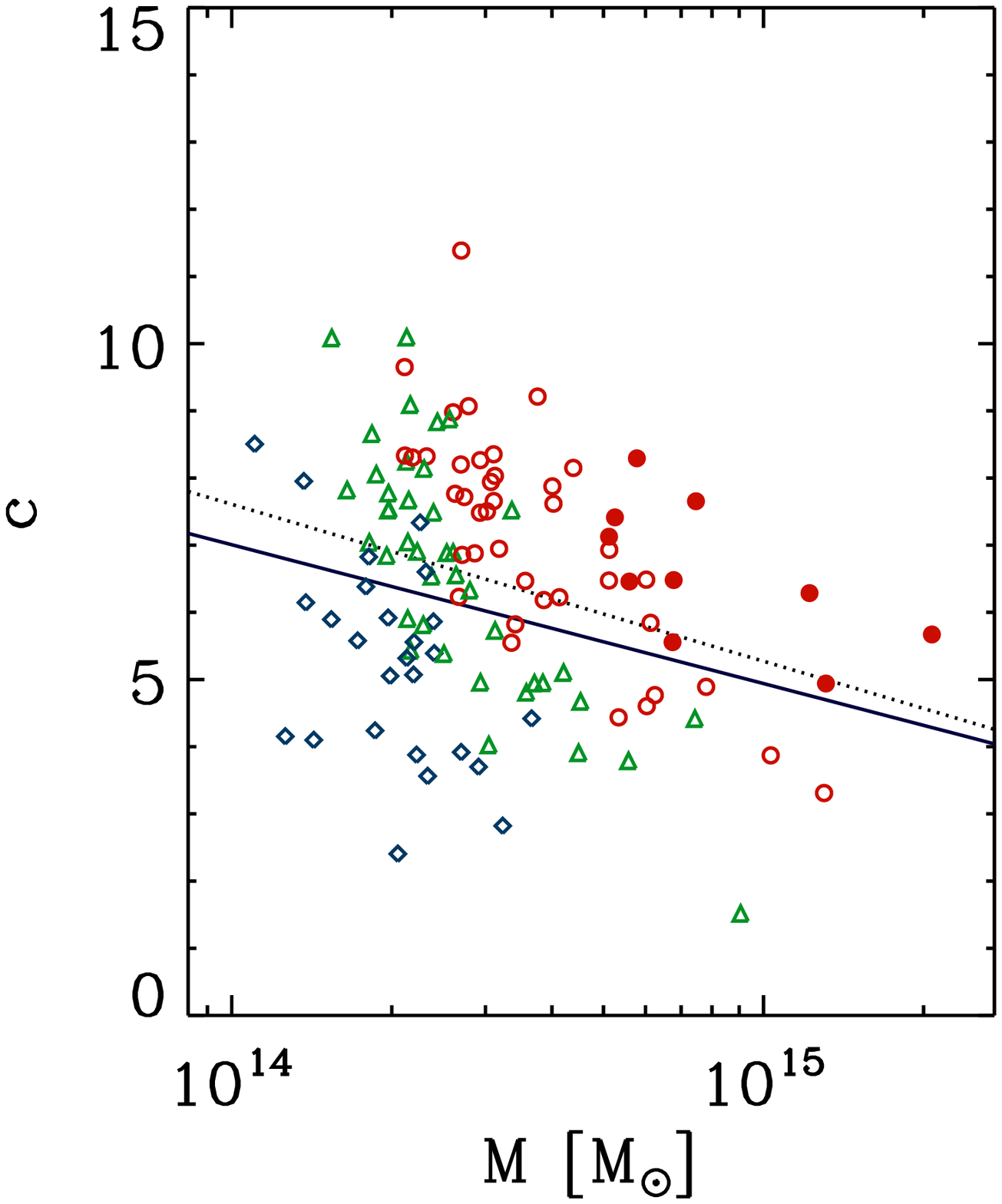,width=\fwid}
}
\caption{The concentration--mass relation for all the clusters
  extracted from cosmological box of
  \protect\cite{2004MNRAS.348.1078B}.  Diamonds, triangles and circles
  are for objects with $\tsl<1.5$ keV, $1.5 < \tsl <2$ keV and $\tsl
  >2$ keV, respectively. The objects used in this paper are indicated
  by filled symbols. In both panels the dotted line represents the
  least--squares fit on the points, the solid line is the model
  proposed by \protect\cite{2001ApJ...554..114E}. The left and the
  right panels correspond to fitting over the whole radial range, out
  to $r_{vir}$, and after excluding the central region within $0.05
  r_{vir}$, respectively. } \label{fi:cm-true}
\end{figure*}

\begin{figure}
\centerline{
\psfig{file=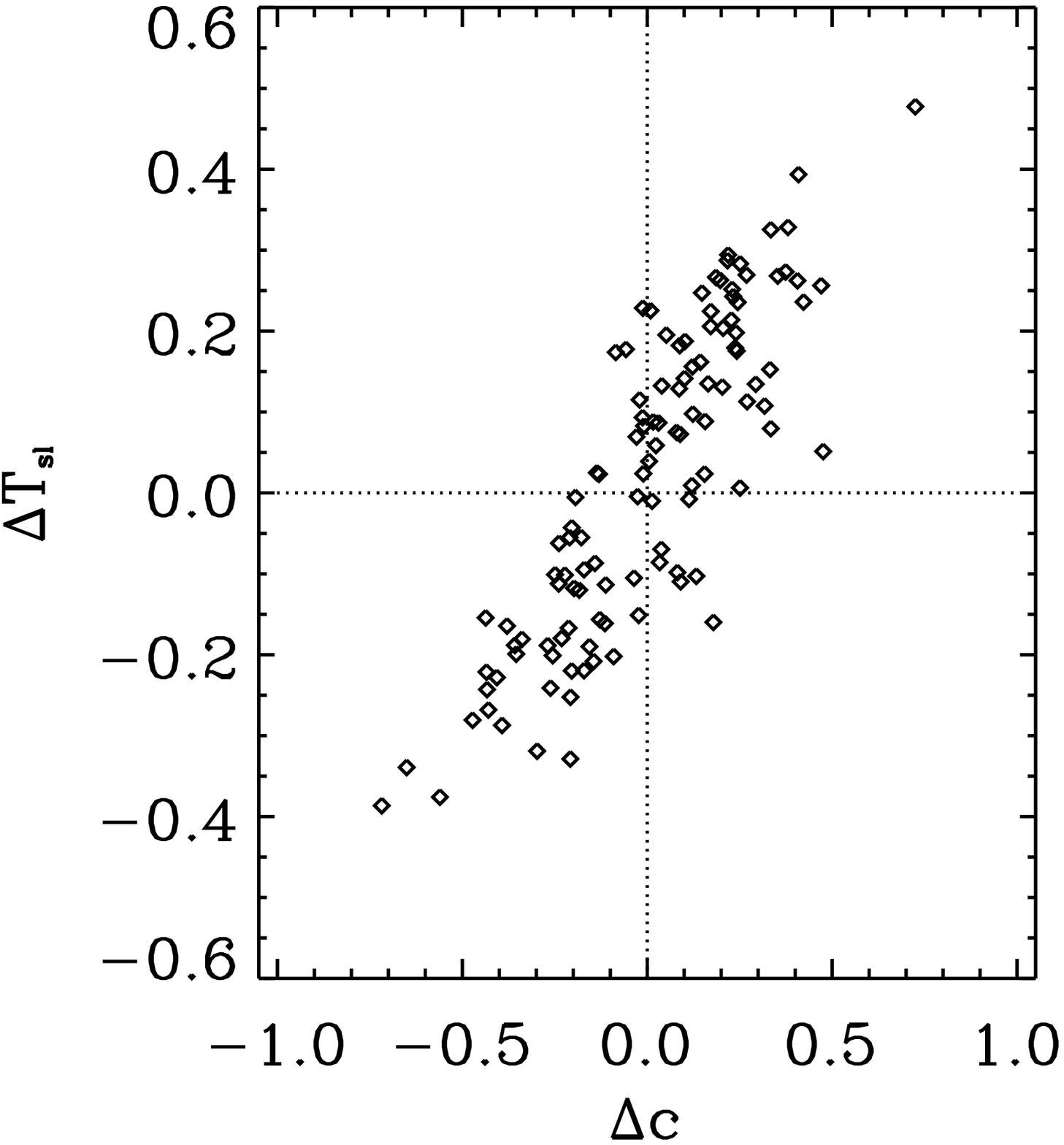,width=\fwid}
}
\caption{The correlation between the scatter in the temperature--mass
  relation and the scatter in the concentration--mass relation. Results
  are shown for the same objects as in Fig. \protect\ref{fi:cm-true}.}
\label{fi:ct-corr}
\end{figure}

In this section, we first study the $c$--$M$ relation on the
three--dimensional mass profiles computed directly from the simulation
data, so that no observational effects are included. Then, we will
show how well this relation can be reconstructed from the analysis of
combined X--ray/SZ observations (using Method 2), for the subset of
clusters having $\tsl \magcir 3$ keV. Since the behaviour of the $c-M$
relation strongly depends on cosmological parameters, we limit our
analysis to those objects belonging to the first simulation (i.e. C1,
C2, C3 and from C5 to C11).  Finally, we also compare our results with
the mass--concentration relation proposed by
\cite{2001ApJ...554..114E}, using the cosmological parameters assumed
in our simulations (see Section 2).

\fig{fi:cm-true} shows the $c$--$M$ relation computed over the 117
clusters identified in the simulation box and having mass assigned by
a friends-of-friends algorithm above $10^{14}\msun$
\citep{2004MNRAS.348.1078B}. The left panel of the figure shows the
result of fitting the integrated mass profile to the NFW formula out
to $r_{vir}$, while in the right panel we exclude the central region
with $r<0.05 r_{vir}$. Note that excluding the inner regions from the
fit has the effect of decreasing the resulting value of $c$, thus
bringing the relation in closer agreement with that by
\cite{2001ApJ...554..114E}, which is based on purely N--body
simulations.  Including gas cooling and star formation leads to the
formation of a stellar component, which generally has a more sharply
peaked mass distribution with respect to the dark matter one.  As a
consequence, one observes a steepening of the total mass profile in
the core, with a subsequent increase of the concentration parameter.
Furthermore, gas cooling is also known to induce adiabatic
contraction, i.e. a steepening of the dark matter profile in reaction
to the strong increase of gas density in the core regions
\citep[e.g.][]{2004ApJ...616...16G}.  We point out that the increase
of $c$ is much reduced when fitting the differential mass profiles,
which gives less weight to the inner halo regions, rather than the
integrated mass profile. However, the case of fitting the integrated
mass is more relevant for \tsz\ and X--ray studies, since the
application of the hydrostatic equilibrium equation gives the
integrated mass profile, rather than the differential one.

In \fig{fi:cm-true}, the colour of each symbol indicates the
spectroscopic--like temperature of the corresponding cluster. We
divide the clusters into 3 temperature bins. We find that hotter
clusters tend to have a slightly larger $c$ than colder ones, at a
fixed mass. The effect can be seen more clearly in
\fig{fi:ct-corr}, which reports the fractional deviation of the
spectroscopic--like temperature, $\Delta\tsl$, and concentration $\Delta
c$ from the mean $\tsl$--$M$ and $c$--$M$ relations respectively. We find
that the correlation factor is quite high, with $r=0.86$. This correlation
is explained by considering that clusters with larger $c$ have a
deeper gravitational potential and thus a hotter ICM. We point out
that this does not depend on the use of the spectroscopic--like
temperature. In fact, when using the mass--weighted temperature,
$\tmw= \sum m_i T_i / \sum m_i$, we find that this
correlation is only slightly weaker, with $r=0.77$, but still significant.
As a consequence, selecting the clusters through their (X--ray)
temperature (as we do in this paper) has the effect of biasing high
the recovered $c$--$M$ relation. 

Our results go in the opposite direction with respect to those by
  \cite{2008arXiv0808.4099Y}, who instead find a negative correlation
  between temperature and concentration in a set of simulated galaxy
  clusters. These authors based their analysis on a cosmological box
  simulated with the Eulerian FLASH code. Tracing the origin of this
  difference between our and their results would require a detailed
  comparison. We note that the run analysed by
  \cite{2008arXiv0808.4099Y} includes quite difference physical
  processes: it assumes gas pre--heating at $z=3$, while it does not
  include radiative cooling, star formation and stellar feedback. We
  also note that the size of the grid in their simulation is
  $250\,h^{-1}$kpc, thus implying that $r_{500}$ in a typical cluster
  is sampled only with few resolution elements.

\begin{figure}
\centerline{
\psfig{file=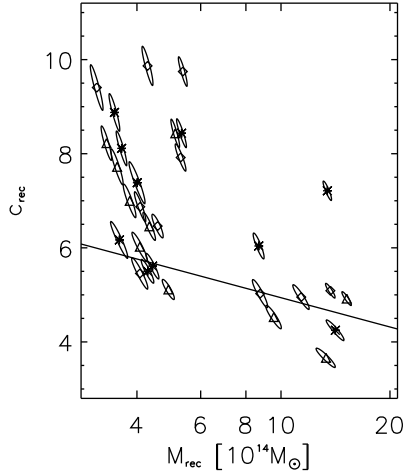,width=\fwid}
}
\caption{The concentration--mass relation recovered from
  deprojection. The solid line represents the model by
  \protect\cite{2001ApJ...554..114E}.  The ellipses show the $1\sigma$
  confidence regions. Diamonds, triangles and stars are for
  the reconstruction from the projection along the $x$, $y$ and $z$
  axis respectively
}
\label{fi:cm-rec}
\end{figure}

\begin{figure}
\psfig{file=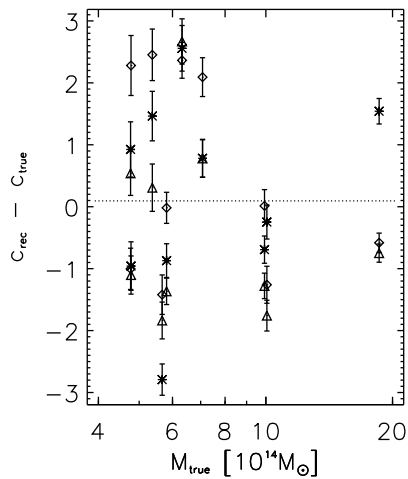,width=\fwid}
\caption{The difference between the recovered and the true
  concentration parameter as a function of the true cluster mass
  within $r_{vir}$. The errorbars show the 1$\sigma$ statistical error
  from the recovery of the cluster mass profile.}\label{fi:conc-err}
\end{figure}

The results discussed in this paragraph and shown in
\figs{fi:cm-rec}{fi:conc-err} have been obtained by applying Method 2
out to $r_{vir}$. The procedure is discussed in detail in Section
\ref{sub:results:met2}.  In \fig{fi:cm-rec} we show the $c$--$M$
relation, as it would be recovered from the deprojection of SZ/X--ray
images.  We find that it is systematically higher than the expected
one. This shift is due to the temperature cut in selecting clusters,
which tends to favor more concentrated systems. Finally, we show in
more detail in \fig{fi:conc-err} our results on the concentration
parameter. We report the error on $c$ as a function of the cluster
mass. We find that $\langle c_{rec}-c_{true} \rangle = 0.1 \pm 1.5$,
with no significant trend with the cluster mass. This result
demonstrates that the value of $c$ is recovered unbiased on average,
although with a fairly large scatter.

\section{Conclusions} 
\label{sec:total-concl} 
Correctly measuring the total collapsed mass of galaxy clusters is of
fundamental importance for these objects to be used as tools for
precision cosmology.  In cluster studies based on the observations of
the intra--cluster medium (ICM) the mass is inferred by applying the
equation of hydrostatic equilibrium (HE) under the assumption of
spherical symmetry, thus requiring that both the gas density and the
temperature profiles are measured to good accuracy. The observational
analyses carried out to so far are essentially based on X--ray
data. Although X--ray surface imaging provides robust measurements of
the gas density profiles out to a significant fraction of the cluster
virialized regions, temperature measurements from spatially resolved
X--ray spectroscopy are in general much noisier and restricted to
smaller radii.  In this paper we addressed the issue of measuring
cluster masses by using a combination of Sunyaev--Zeldovich (SZ) and
X--ray imaging, thereby avoiding X--ray spectroscopy.

To this purpose we presented in this paper a development of the
maximum--likelihood deprojection technique described by
\cite{2007MNRAS.382..397A}, in which we implemented the solution of
the hydrostatic equilibrium equation, so as to derive profiles of gas
density, temperature and total mass simultaneously. We applied this
method to cosmological hydrodynamical simulations of galaxy clusters.
After quantifying the intrinsic level of HE violation in our
simulations, we applied two different methods to recover the total
mass profile from the deprojection of SZ and X--ray maps, both methods
being based on the assumption of HE. Method 1 adopts a
model--independent approach, in which the fitting parameters are
represented by the values of the gas density and total mass profiles
within 15 radial bins, with the only constraint of increasing mass
with radius. Method 2 is instead based on assuming the mass profile by
\cite{NA97.1} (NFW). In this case, the mass profile is characterized
by only 2 fitting parameters, namely the concentration $c$ and the
total mass $M$.

The main results of our analysis be summarized as follows.
\begin{itemize}
\item In keeping with previous analyses
  \citep[e.g.][]{2004MNRAS.351..237R, 2004MNRAS.355.1091K,
  2006MNRAS.369.2013R,2007ApJ...655...98N,2008arXiv0808.1111P}, we
  find that deviations from hydrostatic equilibrium are quite common
  in simulated clusters, due to the presence of pressure
  support associated to stochastic and bulk gas motions. These
  deviations lead to an average mass underestimate of about 10 per
  cent within $r_{500}$. It increases at larger radii, reaching about
  20 per cent at $r_{vir}$, due to the presence of ongoing mergers and
  continuous gas accretion, which characterize the cluster outskirts.
\item From Method 1, we find that in the inner regions (out to about
  $r=0.15 r_{vir}$) the mass is recovered with an underestimate of
  about 10--15 per cent, largely due to the violation of hydrostatic
  equilibrium. On the other hand, the reconstructed mass becomes
  closer to the true one when moving to the outskirts. In this case,
  the effect of noise in the deprojection algorithm tends to
  compensate the underestimate due to the violation of the hydrostatic
  equilibrium. Generally, the fairly large number of fitting
  parameters causes a significant scatter (about 15 per cent) in the
  recovered mass profiles.
\item As for the Method 2, we find that the total mass is
  underestimated by a larger amount, about 20 per cent, when the
  deprojection is performed from $r_{vir}$. Indeed, using a fixed NFW
  functional form for the mass profile forces the mass underestimate
  found at the outermost radius to propagate to smaller
  radii. Therefore, the overall lower normalization of the mass
  profile is just the consequence of the larger violation of the HE
  found at $r_{vir}$. On the other hand, assuming a fixed functional
  form for the mass profile leads to a more stable, although more
  biased, reconstruction, with a scatter of about 10 per cent, thus
  lower than found with Method 1.
\item As expected, fitting the model profiles in the deprojection
  within $r_{500}$ and extrapolating out to the virial radius is a
  safer procedure, thus in agreement with the results by
  \cite{2008arXiv0808.1111P}. The mass underestimate is reduced to
  $\sim 10$ percent, but at the expense of increasing the
  cluster-to-cluster scatter to about 20 per cent.
\item We verified that the choice of the line of sight generally
  affects the mass reconstruction. With Method 1, we show that the
  mass reconstructed from the projection performed along the maximum
  elongation axis is generally larger, by about 10 per cent, than from
  the other two projection directions. Using instead Method 2, we find
  that the typical scatter between the masses reconstructed from
  different projections of the same cluster is typically of only a few
  percent.
\item The relation between concentration parameter of the NFW density
  profile and cluster mass shows a strong correlation with the
  temperature of the cluster: at a fixed mass, hotter clusters tend to
  be more concentrate. For this reason, in our selection of clusters
  having $\tsl \ge 3$~keV we find a $c-M$ relation having higher
  normalization than that calibrated from N--body simulations
  \citep[e.g.,][]{2001ApJ...554..114E}. 
\item Using Method 2, the concentration parameter of the NFW profile
  is recovered on average without any significant bias, but with a
  significant scatter of $\Delta c\simeq 1.5$.
\end{itemize}

Our results lend support to the efficiency of combining X--ray and tSZ
imaging data to recover the total mass profiles of galaxy clusters. In
fact, the main bias that we found is intrinsic, since it is due to
deviations from perfect hydrostatic equilibrium.  This approach has
several advantages with respect to the traditional one based on X--ray
spectroscopy. Firstly, the temperature recovered from the fit of the
X--ray spectra generally differs from the electron temperature, by an
amount which depends on the degree of complexity of the ICM thermal
structure \citep[e.g., ][]{2004MNRAS.354...10M,
  2006ApJ...640..710V}. Secondly, X--ray surface brightness profiles
can be obtained with good precision with a relatively small number of
photon counts ($\sim 10^2$),
while at least ten times more photons are required for a
  reliable spatially resolved spectroscopy. 
 Also, once the cosmic and instrumental backgrounds are under
control, this opens the possibility of tracing the surface brightness
over a large portion of the cluster virial regions, as already
demonstrated with ROSAT-PSPC imaging data
\citep[e.g.,][]{1999ApJ...525...47V,2005A&A...439..465N}. Since the
tSZ has the potential of covering a large range in gas density, its
combination with low--background X--ray imaging data would allow one
to better characterize the outskirts of galaxy clusters. Moreover,
another advantage is that the tSZ is independent of redshift, so its
combination with X--ray surface brightness will substantially improve
the analysis of cluster properties at high redshift, where accurate
X--ray spectroscopy may be very hard, if not impossible, to obtain.

A limitation of the analysis presented in this paper is that we did
not include realistic backgrounds in the generation of the X--ray and
tSZ maps. As we have just mentioned, there are interesting
perspectives for a good characterization of the X--ray
background. However, the situation could be more complicated for the
tSZ background. In this case, contaminating signals from unresolved
point--like radio sources \citep[e.g.,][]{2006A&A...447..405B} and
fore/background galaxy groups \citep[e.g.,][]{2007arXiv0704.2607H}
could affect the tSZ signal in the cluster outskirts. 
In this respect, the possibility of performing multi--frequency
observations with the good angular resolution offered by
  interferometric techniques will surely help in characterizing and
removing these contaminations.

Single--dish sub-millimetric telescopes of the next generation promise
to provide tSZ images of clusters with a spatial resolution of few
tens of arcsec, while covering fairly large field of views, with
10--20 arcmin aside, with a good sensitivity. At the same time, future
satellites for X--ray surveys (e.g. eROSITA) will have the capability
of surveying large areas of the sky with high sensitivity and good
control of the background. These observational facilities will open
the possibility of carrying out in survey mode high--quality tSZ and
X--ray imaging for a large number of clusters. The application of
deprojection methods, like those presented in this paper will provide
reliable determinations of both the gas mass and total mass
profiles. This will greatly help to fully exploit the potentiality of
galaxy clusters as tools for precision cosmology.

\section*{Acknowledgments.} 
We thank Giuseppe Murante and Elena Rasia for useful discussions.  The
simulations have been carried out at the ``Centro Interuniversitario
del Nord-Est per il Calcolo Elettronico'' (CINECA, Bologna), with CPU
time assigned thanks to an INAF--CINECA grant and to an agreement
between CINECA and the University of Trieste. This work has been
partially supported by the INFN PD-51 grant, by the INAF-PRIN06 Grant
and by a ASI--AAE grant. We acknowledge the financial contribution
from contracts ASI-INAF I/023/05/0 and I/088/06/0.

\bibliographystyle{mn2e}

\bibliography{master}

\clearpage

\end{document}